\documentclass[english]{IEEEtran}
\usepackage[T1]{fontenc}
\usepackage[latin9]{inputenc}
\usepackage{array}
\usepackage{verbatim}
\usepackage{amsmath}
\usepackage{graphicx}
\usepackage{amssymb}

\makeatletter

\providecommand{\tabularnewline}{\\}

\usepackage{amsthm}
\usepackage{dsfont}
\usepackage{array}
\usepackage{mathrsfs}
\usepackage{cite}
\usepackage{comment}

\usepackage{babel}

\makeatother

\usepackage{babel}

\makeatother

\usepackage{babel}

\makeatother

\usepackage{babel}

\begin{document}
\bibliographystyle{IEEEtran}

\title{On the Role of Mobility for Multi-message Gossip}

\author{Yuxin Chen, Sanjay Shakkottai and Jeffrey G. Andrews %
\thanks{Y. Chen is with the Department of Electrical Engineering and the Department
of Statistics, Stanford University, Stanford CA 94305, USA (email:
yxchen@stanford.edu). S. Shakkottai is with the Department of Electrical
and Computer Engineering, the University of Texas at Austin, Austin
TX 78712, USA (email: shakkott@ece.utexas.edu). J. G. Andrews is with
the Department of Electrical and Computer Engineering, the University
of Texas at Austin, Austin TX 78712, USA (email: jandrews@ece.utexas.edu).
The contact author is S. Shakkottai. 

This research has been supported by the DARPA Information Theory for
Mobile Ad Hoc Networks (IT-MANET) program and NSF Grant 0721380. It
has been presented in part at the IEEE Infocom 2011 \cite{ChenInfocom}.
Manuscript date: \today.%
}}

\maketitle
\theoremstyle{plain}\newtheorem{lem}{Lemma}\newtheorem{theorem}{Theorem}\newtheorem{corollary}{Corollary}

\theoremstyle{definition}\newtheorem{remark}{Remark}
\begin{abstract}
We consider information dissemination in a large $n$-user wireless
network in which $k$ users wish to share a unique message with all
other users. Each of the $n$ users only has knowledge of its own
contents and state information; this corresponds to a one-sided push-only
scenario. The goal is to disseminate all messages efficiently, hopefully
achieving an order-optimal spreading rate over unicast wireless random
networks. First, we show that a random-push strategy -- where a user
sends its own or a received packet at random -- is order-wise suboptimal
in a random geometric graph: specifically, $\Omega\left(\sqrt{n}\right)$
times slower than optimal spreading. It is known that this gap can
be closed if each user has ``full'' mobility, since this effectively
creates a complete graph. We instead consider velocity-constrained
mobility where at each time slot the user moves locally using a discrete
random walk with velocity $v(n)$ that is much lower than full mobility.
We propose a simple two-stage dissemination strategy that alternates
between individual message flooding (``self promotion'') and random
gossiping. We prove that this scheme achieves a close to optimal spreading
rate (within only a logarithmic gap) as long as the velocity is at
least $v(n)=\omega(\sqrt{\log n/k})$. The key insight is that the
mixing property introduced by the partial mobility helps users to
spread in space within a relatively short period compared to the optimal
spreading time, which macroscopically mimics message dissemination
over a complete graph.

{}
\end{abstract}

\section{Introduction}

In wireless ad hoc or social networks, a variety of scenarios require
agents to share their individual information or resources with each
other for mutual benefits. A partial list includes file sharing and
rumor spreading \cite{QiuSri04,YanDeV04,Bor87,KarSchSheVoc00}, distributed
computation and parameter estimation \cite{Tsi84,KasSri07,BoydShah06,NedOzd09,JunShaShi10},
and scheduling and control \cite{ModShaZus,EryOzdShaMod10}. Due to
the huge centralization overhead and unpredictable dynamics in large
networks, it is usually more practical to disseminate information
and exchange messages in a decentralized and asynchronous manner to
combat unpredictable topology changes and the lack of global state
information. This motivates the exploration of dissemination strategies
that are inherently simple, distributed and asynchronous while achieving
optimal spreading rates.

\subsection{Motivation and Related Work}

Among distributed asynchronous algorithms, gossip algorithms are a
class of protocols which propagate messages according to rumor-style
rules, initially proposed in \cite{Dem88}. Specifically, suppose
that there are $k\leq n$ distinct pieces of messages that need to
be flooded to all $n$ users in the network. Each agent in each round
attempts to communicate with one of its neighbors in a random fashion
to disseminate a limited number of messages. There are two types of
push-based strategies on selecting which message to be sent in each
round: (a) one-sided protocols that are based only on the disseminator's
own current state; and (b) two-sided protocols based on current states
of both the sender and the receiver. Encouragingly, a simple uncoded
one-sided push-only gossip algorithm with random message selection
and peer selection is sufficient for efficient dissemination in some
cases like a static complete graph, which achieves a spreading time
of $\Theta\left(k\log n\right)$ %
\footnote{The standard notion $f(n)=\omega\left(g(n)\right)$ means $\underset{n\rightarrow\infty}{\lim}g(n)/f(n)=0$;
$f(n)=o\left(g(n)\right)$ means $\underset{n\rightarrow\infty}{\lim}f(n)/g(n)=0$;
$f(n)=\Omega\left(g(n)\right)$ means $\exists$ a constant $c$ such
that $f(n)\geq cg(n)$; $f(n)=O\left(g(n)\right)$ means $\exists$
a constant $c$ such that $f(n)\leq cg(n)$; $f(n)=\Theta\left(g(n)\right)$
means $\exists$ constants $c_{1}$ and $c_{2}$ such that $c_{1}g(n)\leq f(n)\leq c_{2}g(n)$.%
}, within only a logarithmic gap with respect to the optimal lower
limit $\Theta(k)$ \cite{Pittel1987,GossipTutorialShah,SanHajMas07}.
This type of one-sided gossiping has the advantages of being easily
implementable and inherently distributed.

%
{}

Since each user can receive at most one message in any single slot,
it is desirable for a protocol to achieve close to the fastest possible
spreading time $\Theta\left(k\right)$ (e.g. within a $\text{polylog}(n)$
factor). It has been pointed out, however, that the spreading rate
of one-sided random gossip algorithms is frequently constrained by
the network geometry, e.g. the conductance of the graph \cite{MoskSha08,GossipTutorialShah}.
For instance, for one-sided rumor-style all-to-all spreading (i.e.
$k=n$), the completion time $T$ is much lower in a complete graph
$\left(T=O\left(n\log n\right)\right)$ than in a ring $\left(T=\Omega(n^{2})\right)$.
Intuitively, since each user can only communicate with its nearest
neighbors, the geometric constraints in these graphs limit the location
distribution of all copies of each message during the evolution process,
which largely limits how fast the information can flow across the
network. In fact, for message spreading over static wireless networks,
one-sided uncoded push-based random gossiping can be quite inefficient:
specifically up to $\Omega\left(\sqrt{\frac{n}{\text{poly}(\log n)}}\right)$
times slower than the optimal lower limit $\Theta\left(k\right)$
(i.e. a polynomial factor away from the lower bound), as will be shown
in Theorem \ref{thm-Random-Push-Static}.

{}

Although one-sided random gossiping is not efficient for static wireless
networks, it may potentially achieve better performance if each user
has some degree of mobility -- an intrinsic feature of many wireless
and social networks. For instance, full mobility%
\footnote{By full mobility, we mean that the location of the mobile is distributed
independently and uniformly random over the entire network over consecutive
time-steps (i.e., the velocity of the mobile can be {}``arbitrarily
large''). This is sometimes also referred to in literature as the
i.i.d. mobility model. In this paper, we consider nodes with {}``velocity-limited''
mobility capability.%
} changes the geometric graph with transmission range $O\left(\sqrt{\frac{\log n}{n}}\right)$
to a complete graph in the unicast scenario. Since random gossiping
achieves a spreading time of $\Theta\left(n\text{log}(n)\right)$
for \emph{all-to-all} spreading over a complete graph \cite{SanHajMas07,GossipTutorialShah},
this allows near-optimal spreading time to be achieved within a logarithmic
factor from the fundamental lower limit $\Theta\left(n\right)$. However,
how much benefit can be obtained from more realistic mobility -- which
may be significantly lower than idealized best-case full mobility
-- is not clear. Most existing results on uncoded random gossiping
center on evolutions associated with static homogeneous graph structure
or a fixed adjacency matrix, which cannot be readily extended for
dynamic topology changes. To the best of our knowledge, the first
work to analyze gossiping with mobility was \cite{MobilityGossip},
which focused on \textit{energy-efficient} distributed averaging instead
of \textit{time-efficient} message propagation. Another line of work
by Clementi \emph{et. al.} investigate the speed limit for information
flooding over Markovian evolving graphs (e.g. \cite{Clementi2011,Baumann2009,Clementi2012}),
but they did not study the spreading rate under multi-message gossip.
Recently, Pettarin \emph{et. al.} \cite{Pettarin2011} explored the
information spreading over sparse mobile networks with no connected
components of size $\Omega\left(\log n\right)$, which does not account
for the dense (interference-limited) network model we consider in
this paper.

For a broad class of graphs that include both static and dynamic graphs,
the lower limit on the spreading time can be achieved through random
linear coding where a random combination of all messages are transmitted
instead of a specific message \cite{DebMedCho06}, or by employing
a two-sided protocol which always disseminates an innovative message
if possible \cite{SanHajMas07}. Specifically, through a unified framework
based on dual-space analysis, recent work \cite{Hae2011} demonstrated
that the optimal all-to-all spreading time $\Theta(n)$ can be achieved
for a large class of graphs \cite{Hae2011} including complete graphs,
geometric graphs, and the results hold in these network models even
when the topology is allowed to change dynamically at each time. However,
performing random network coding incurs very large computation overhead
for each user, and is not always feasible in practice. On the other
hand, two-sided protocols inherently require additional feedback that
increases communication overhead. Also, the state information of the
target may sometimes be unobtainable due to privacy or security concerns.
Furthermore, if there are $k\ll\sqrt{n}$ messages that need to be
disseminated over a static uncoordinated unicast wireless network
or a random geometric graph with transmission radius $\Theta\left(\sqrt{\frac{1}{n}\text{poly}\left(\log n\right)}\right)$,
neither network coding nor two-sided protocols can approach the lower
limit of spreading time $\Theta(k)$. This arises due to the fact
that the diameter of the underlying graph with transmission range
$\Theta\left(\sqrt{\frac{\text{poly}\left(\log n\right)}{n}}\right)$
scales as $\Omega\left(\sqrt{\frac{n}{\text{poly}(\log n)}}\right)$,
and hence each message may need to be relayed through $\Omega\left(\sqrt{\frac{n}{\text{poly}(\log n)}}\right)$
hops in order to reach the node farthest to the source.

%
{}

Another line of work has studied spreading scaling laws using more
sophisticated non-gossip schemes over static wireless networks, e.g.
\cite{Zhe2006,SubShaAra}. Recently, Resta \textit{et. al.} \cite{ResSan2010}
began investigating broadcast schemes for mobile networks with a \textit{single
static} source constantly propagating new data, while we focus on
a different problem with multiple mobile sources each sharing distinct
message. Besides, \cite{ResSan2010} analyzed how to combat the adverse
effect of mobility to ensure the same pipelined broadcasting as in
static networks, whereas we are interested in how to take advantage
of mobility to overcome the geometric constraints. In fact, with the
help of mobility, simply performing random gossiping -- which is simpler
than most non-gossip schemes and does not require additional overhead
-- is sufficient to achieve optimality. 

Finally, we note that gossip algorithms have also been employed and
analyzed for other scenarios like distributed averaging, where each
node is willing to compute the average of all initial values given
at all nodes in a decentralized manner, e.g. \cite{Tsi84,BoydShah06}.
The objective of such distributed consensus is to minimize the total
number of computations. It turns out that the convergence rates of
both message sharing and distributed averaging are largely dependent
on the eigenvalues or, more specifically, the mixing times of the
graph matrices associated with the network geometry \cite{BoydShah06,GossipTutorialShah}.

\subsection{Problem Definition and Main Modeling Assumptions}

Suppose there are $n$ users randomly located over a square of unit
area. The task is to disseminate $k\leq n$ distinct messages (each
contained in one user initially) among all users. The message spreading
can be categorized into two types: (a) \textit{single-message dissemination}:
a single user (or $\Theta(1)$ users) wishes to flood its message
to all other users; (b) \textit{multi-message dissemination}: a large
number $k$$(k\gg1)$ of users wish to spread individual messages
to all other users. We note that distinct messages may not be injected
into the network simultaneously. They may arrive in the network (possibly
in batches) sequentially, but the arrival time information is unknown
in the network.

Our objective is to design a gossip-style one-sided algorithm in the
absence of coding, such that it can take advantage of the intrinsic
feature of mobility to accelerate dissemination. Only the {}``push''
operation is considered in this paper, i.e. a sender determines which
message to transmit solely based on its own current state, and in
particular not using the intended receiver's state. We are interested
in identifying the range of the degree of mobility within which our
algorithm achieves near-optimal spreading time $O\left(k\text{ poly}\left(\log n\right)\right)$
for each message regardless of message arrival patterns. Specifically,
our MOBILE PUSH protocol achieves a spreading time $O\left(k\log^{2}n\right)$
as stated in Theorem \ref{thm:DiscreteMP} for the mobility that is
significantly lower than the idealized full mobility. As an aside,
it has been shown in \cite{SanHajMas07,DebMedCho06} that with high
probability, the completion time for one-sided uncoded random gossip
protocol over complete graphs is lower bounded by $\Omega\left(k\log n\right)$,
which implies that in general the logarithmic gap from the universal
lower limit $\Theta\left(k\right)$ cannot be closed with uncoded
one-sided random gossiping.

Our basic network model is as follows. Initially, there are $n$ users
uniformly distributed over a unit square. We ignore edge effects so
that every node can be viewed as homogeneous. Our models and analysis
are mainly based on the context of wireless ad hoc networks, but one
can easily apply them to other network scenarios that can be modeled
as a random geometric graph of transmission radius $\Theta\left(\sqrt{\log n/n}\right)$.

\textbf{\textit{Physical-Layer Transmission Model}}. Each transmitter
employs the same amount of power $P$, and the noise power density
is assumed to be $\eta$. The path-loss model is used such that node
$j$ receives the signal from transmitter $i$ with power $Pr_{ij}^{-\alpha}$,
where $r_{ij}$ denotes the Euclidean distance between $i$ and $j$
with $\alpha>2$ being the path loss exponent. Denote by $\mathcal{\mathcal{T}}(t)$
the set of transmitters at time $t$. We assume that a packet from
transmitter $i$ is successfully received by node $j$ at time $t$
if \begin{align}
\text{SINR}_{ij}(t): & =\frac{Pr_{ij}^{-\alpha}}{\eta+\sum\limits _{k\neq i,k\in\mathcal{T}(t)}Pr_{kj}^{-\alpha}}\geq\beta,\end{align}
 where $\text{SINR}_{ij}(t)$ is the signal-to-interference-plus-noise
ratio (SINR) at $j$ at time $t$, and $\beta$ the SINR threshold
required for successful reception. For simplicity, we suppose only
one fixed-size message or packet can be transmitted for each transmission
pair in each time instance.

\begin{table}
\caption{Summary of Notation}

\centering{}\begin{tabular}{l>{\raggedright}p{2.1in}}
$v(n)$  & velocity\tabularnewline
$m$  & the number of subsquares; $m=1/v^{2}(n)$\tabularnewline
$n$  & the number of users/nodes\tabularnewline
$k$  & the number of distinct messages\tabularnewline
$M_{i}$  & the message of source $i$\tabularnewline
$A_{k}$  & subsquare $k$\tabularnewline
$N_{i}(t)$,$\mathcal{N}_{i}(t)$  & the number, and the set of nodes containing $M_{i}$ at time $t$\tabularnewline
$N_{i,A_{k}}(t)$,$\mathcal{N}_{i,A_{k}}(t)$  & the number, and the set of nodes containing $M_{i}$ at subsquare
$A_{k}$ at time $t$\tabularnewline
$S_{i}(t),\mathcal{S}_{i}(t)$  & the number, and the set of messages node $i$ has at time $t$\tabularnewline
$\alpha$  & path loss exponent\tabularnewline
$\beta$  & SINR requirement for single-hop success\tabularnewline
\end{tabular}%
\end{table}

Suppose that each node can move with velocity $v(n)$ in this mobile
network. We provide a precise description of the mobility pattern
as follows.

\begin{figure}[htbp]
\begin{centering}
\textsf{\includegraphics[scale=0.45]{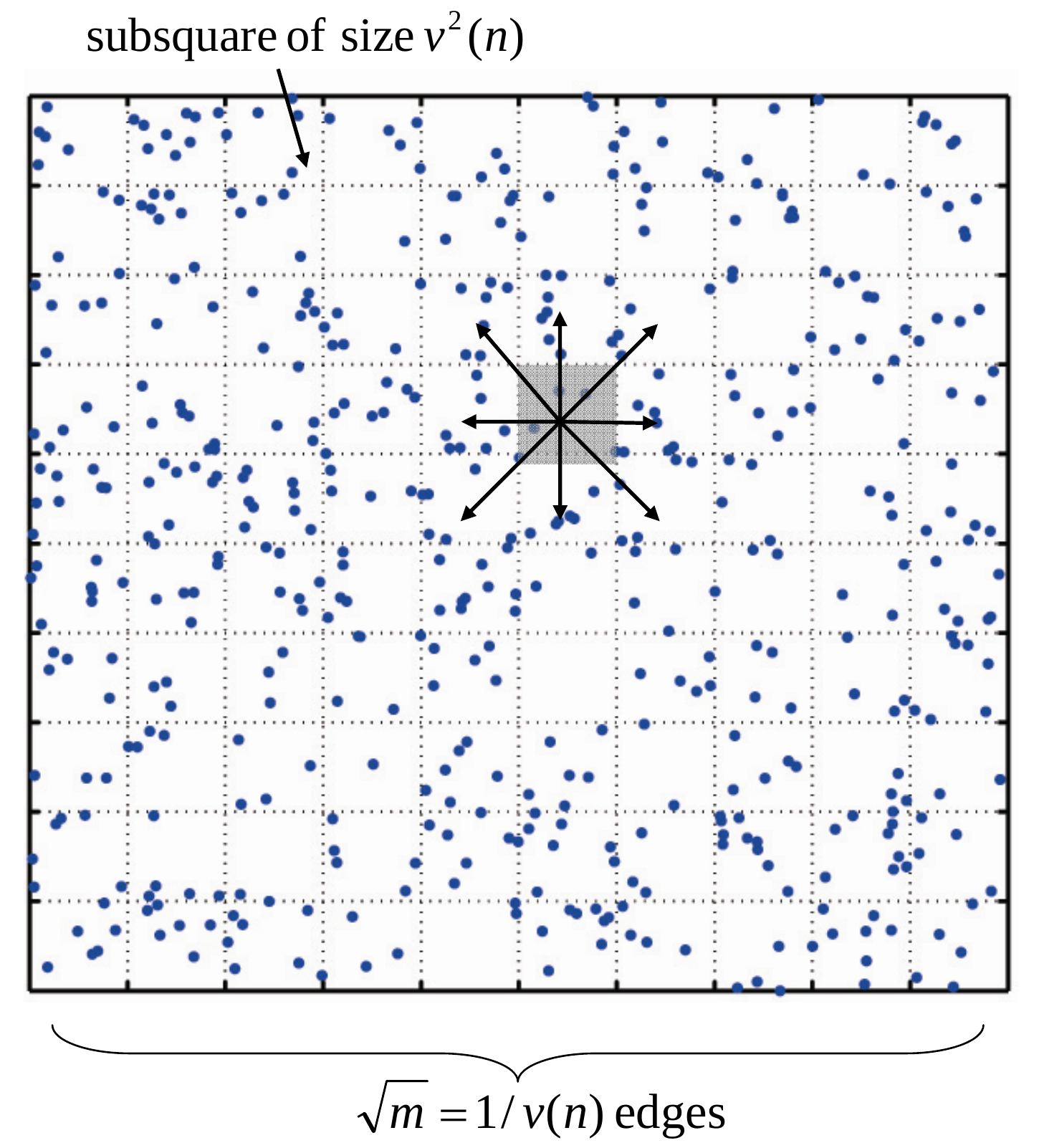}} 
\par\end{centering}

\caption{\label{fig:UnitSquareGrid}The unit square is equally divided into
$m=1/v^{2}(n)$ subsquares. Each node can jump to one of its $8$
neighboring subsquares or stay in its current subsquare with equal
probability $1/9$ at the beginning of each slot.}
\end{figure}

\textbf{\textit{Mobility Model}}. We use a mobility pattern similar
to \cite[Section VIII]{NeeMod05}, which ensures that at steady state,
each user lies in each subsquare with equal probability. Specifically,
we divide the entire square into $m:=1/v^{2}(n)$ subsquares each
of area $v^{2}(n)$ (where $v(n)$ denotes the velocity of the mobile
nodes), which forms a $\sqrt{m}\times\sqrt{m}$ discrete torus. At
each time instance, every node moves according to a \textit{random
walk} on the $\sqrt{m}\times\sqrt{m}$ discrete torus. More precisely,
if a node resides in a subsquare $(i,j)\in\left\{ 1,\cdots,\sqrt{m}\right\} ^{2}$
at time $t$, it may choose to stay in $(i,j)$ or move to any of
the eight adjacent subsquares each with probability $1/9$ at time
$t+1$. If a node is on the edge and is selected to move in an infeasible
direction, then it stays in its current subsquare. The position inside
the new subsquare is selected \textit{uniformly at random}. See Fig.
\ref{fig:UnitSquareGrid} for an illustration.

We note that when $v(n)=1/3=\Theta(1)$, the pattern reverts to the
full mobility model. In this random-walk model, each node moves independently
according to a uniform ergodic distribution. In fact, a variety of
mobility patterns have been proposed to model mobile networks, including
i.i.d. (full) mobility \cite{MobilityCapacityTse}, random walk (discrete-time)
model \cite{GamMamProSha06,YinYanSri08}, and Brownian motion (continuous-time)
pattern \cite{LinMazShr06}. For simplicity, we model it as a discrete-time
random walk pattern, since it already captures intrinsic features
of mobile networks like uncontrolled placement and movement of nodes.

\subsection{Contributions and Organization}

The main contributions of this paper include the following. 
\begin{enumerate}
\item \textbf{Single-message dissemination over mobile networks.} We derive
an upper bound on the single-message ($k=\Theta\left(1\right)$) spreading
time using push-only random gossiping (called RANDOM PUSH) in mobile
networks. A gain of $\Omega\left(v(n)\sqrt{n}/\log^{2}n)\right)$
in the spreading rate can be obtained compared with static networks,
which is, however, still limited by the underlying geometry unless
there is full mobility. 
\item \textbf{Multi-message dissemination over static networks.} We develop
a lower bound on the multi-message spreading time under RANDOM PUSH
protocol over static networks. It turns out that there may exist a
gap as large as $\Omega\left(\frac{\sqrt{n}}{\text{poly}\left(\log n\right)}\right)$
between its spreading time and the optimal lower limit $\Theta\left(k\right)$.
The key intuition is that the copies of each message $M_{i}$ tend
to cluster around the source $i$ at all time instances, which results
in capacity loss. This inherently constrains how fast the information
can flow across the network. 
\item \textbf{Multi-message dissemination over mobile networks.} We design
a \textit{one-sided uncoded} message-selection strategy called MOBILE
PUSH that accelerates multi-message spreading ($k=\omega\left(\log n\right)$)
with mobility. An upper bound on the spreading time is derived, which
is the main result of this paper. Once $v(n)=\omega\left(\sqrt{\frac{\log n}{k}}\right)$
(which is still significantly smaller than full mobility), the near-optimal
spreading time $O\left(k\log^{2}n\right)$ can be achieved with high
probability. The underlying intuition is that if the mixing time arising
from the mobility model is smaller than the optimal spreading time,
the mixing property approximately \textit{uniformizes} the location
of all copies of each message, which allows the evolution to mimic
the propagation over a complete graph. 
\end{enumerate}
The remainder of this paper is organized as follows. In Section \ref{sec:Strategies-and-Main-Results},
we describe our unicast physical-layer transmission strategy and two
types of message selection strategies, including RANDOM PUSH and MOBILE
PUSH. Our main theorems are stated in Section \ref{sec:Strategies-and-Main-Results}
as well, with proof ideas illustrated in Section \ref{thm:DiscreteMP}.
Detailed derivation of auxiliary lemmas are deferred to the Appendix.

\section{Strategies and Main Results\label{sec:Strategies-and-Main-Results}}

The strategies and main results of this work are outlined in this
section, where only the unicast scenario is considered. The dissemination
protocols for wireless networks are a class of scheduling algorithms
that can be decoupled into (a) physical-layer \textit{transmission}
strategies (link scheduling) and (b) message selection strategies
(message scheduling).

One physical-layer transmission strategy and two message selection
strategies are described separately, along with the order-wise performance
bounds.

\subsection{Strategies}

\subsubsection{Physical-Layer Transmission Strategy}

In order to achieve efficient spreading, it is natural to resort to
a decentralized transmission strategy that supports the order-wise
largest number (i.e. $\Theta(n)$) of concurrent successful transmissions
per time instance. The following strategy is a candidate that achieves
this objective with local communication. \vspace{6pt}

\framebox{%
\begin{minipage}[t]{3.2in}%
UNICAST Physical-Layer Transmission Strategy: 
\begin{itemize}
\item At each time slot, each node $i$ is designated as a sender independently
with constant probability $\theta$, and a potential receiver otherwise.
Here, $\theta<0.5$ is independent of $n$ and $k$. 
\item Every sender $i$ attempts to transmit one message to its \textit{nearest}
potential receiver $j(i)$. 
\end{itemize}
\end{minipage}}

\vspace{8pt}
 This simple {}``link'' scheduling strategy, when combined with
appropriate push-based message selection strategies, leads to the
near-optimal performance in this paper. We note that the authors in
\cite{MobilityCapacityTse}, by adopting a slightly different strategy
in which $\theta n$ nodes are randomly designated as senders (as
opposed to link-by-link random selection as in our paper), have shown
that the success probability for each unicast pair is a constant.
Using the same proof as for \cite[Theorem III-5]{MobilityCapacityTse},
we can see (which we omit here) that there exists a constant $c$
such that\begin{equation}
\mathbb{P}\left(\text{SINR}_{i,j(i)}(t)>\beta\right)\geq c\end{equation}
holds for our strategy. Here, $c$ is a constant irrespective of $n$,
but may depend on other salient parameters $P$, $\alpha$ and $\eta$.
That said, $\Theta\left(n\right)$ concurrent transmissions can be
successful, which is order-optimal. For ease of analysis and exposition,
we further assume that physical-layer success events are \textit{temporally
independent} for simplicity of analysis and exposition. Indeed, even
accounting for the correlation yields the same scaling results, detailed
in Remark 1. 

\begin{remark}In fact, the physical-layer success events are correlated
across different time slots due to our mobility model and transmission
strategy. However, we observe that our analysis framework would only
require that the transmission success probability at time $t+1$ is
always a constant irrespective of $n$ given the node locations at
time $t$. To address this concern, we show in Lemma \ref{lemmaConcentration-3}
that for any $m<n/\left(32\log n\right)$, the number of nodes $N_{A_{i}}$
residing in each subsquare $A_{i}$ is bounded within $\left[\frac{n}{6m},\frac{7n}{3m}\right]$
$ $with probability at least $1-2n^{-3}$. Conditional on this high-probability
event that $N_{A_{i}}\in\left[\frac{n}{6m},\frac{7n}{3m}\right]$
with all nodes in each subsquare uniformly located, we can use the
same proof as \cite[Theorem III-5]{MobilityCapacityTse} to show that
$\mathbb{P}\left(\text{SINR}_{i,j(i)}(t)>\beta\right)\geq c$ holds
for some constant $c$. \end{remark}

Although this physical-layer transmission strategy supports $\Theta\left(n\right)$
concurrent local transmissions, it does not tell us how to take advantage
of these resources to allow efficient propagation. This will be specified
by the message-selection strategy, which potentially determines how
each message is propagated and forwarded over the entire network.

\subsubsection{Message Selection Strategy}

We now turn to the objective of designing a one-sided message-selection
strategy (only based on the transmitter's current state) that is efficient
in the absence of network coding. We are interested in a decentralized
strategy in which no user has prior information on the number of distinct
messages existing in the network. One common strategy within this
class is:

\vspace{6pt}

\framebox{%
\begin{minipage}[t]{3.2in}%
\textbf{RANDOM PUSH} Message Selection Strategy: 
\begin{itemize}
\item In every time slot: each sender $i$ randomly selects one of the messages
it possesses for transmission. 
\end{itemize}
\end{minipage}}

\vspace{8pt}

This is a simple gossip algorithm solely based on random message selection,
which is surprisingly efficient in many cases like a complete graph.
It will be shown later, however, that this simple strategy is inefficient
in a static unicast wireless network or a  random geometric graph
with transmission range $\Theta\left(\sqrt{\frac{\log n}{n}}\right)$.

In order to take advantage of the mobility, we propose the following
alternating strategy within this class:

\vspace{6pt}

\framebox{%
\begin{minipage}[t]{3.2in}%
\textbf{MOBILE PUSH} Message Selection Strategy: 
\begin{itemize}
\item Denote by $M_{i}$ the message that source $i$ wants to spread, i.e.
its own message. 
\item In every odd time slot: for each sender $i$, if it has an individual
message $M_{i}$, then $i$ selects $M_{i}$ for transmission; otherwise
$i$ randomly selects one of the messages it possesses for transmission. 
\item In every even time slot: each sender $i$ randomly selects one of
the messages it has received for transmission. 
\end{itemize}
\end{minipage}}

\vspace{8pt}
 In the above strategy, each sender alternates between random gossiping
and self promotion. This alternating operation is crucial if we do
not know \textit{a priori} the number of distinct messages. Basically,
random gossiping enables rapid spreading by taking advantage of all
available throughput, and provides a non-degenerate approach that
ensures an approximately {}``uniform'' evolution for all distinct
messages. On the other hand, individual message flooding step plays
the role of self-advocating, which guarantees that a sufficiently
large number of copies of each message can be forwarded with the assistance
of mobility (which is not true in static networks). This is critical
at the initial stage of the evolution.

\subsection{Main Results (without proof)}

Now we proceed to state our main theorems, each of which characterizes
the performance for one distinct scenario. Detailed analysis is deferred
to Section \ref{sec:Discrete-Jump-Model}.

\subsubsection{Single-Message Dissemination in Mobile Networks with RANDOM PUSH}

The first theorem states the limited benefits of mobility on the spreading
rate for single-message spreading when RANDOM PUSH is employed. We
note that MOBILE PUSH reverts to RANDOM PUSH for single-message dissemination,
and hence has the same spreading time. 

\begin{theorem}\label{thm:DiscreteSP}Assume that the velocity obeys
$v(n)>\sqrt{\frac{32\log n}{n}}$, and that the number of distinct
messages obeys $k=\Theta\left(1\right)$. RANDOM PUSH message selection
strategy is assumed to be employed in the unicast scenario. Denote
by $T_{\mathrm{sp}}^{\mathrm{uc}}(i)$ the time taken for all users
to receive message $M_{i}$ after $M_{i}$ is injected into the network,
then with probability at least $1-n^{-2}$ we have\begin{equation}
\forall i,\quad T_{\mathrm{sp}}^{\mathrm{uc}}(i)=O\left(\frac{\log n}{v(n)}\right)\quad\text{and}\quad T_{\mathrm{sp}}^{\mathrm{uc}}(i)=\Omega\left(\frac{1}{v(n)}\right).\end{equation}

\end{theorem}

Since the single-message flooding time is $\Omega(\sqrt{n}/\log n)$
under RANDOM PUSH over static wireless networks or random geometric
graphs of radius $\Theta\left(\sqrt{\log n/n}\right)$ \cite{MoskSha08},
the gain in dissemination rate due to mobility is $\Omega\left(v(n)\sqrt{n}/\log^{2}n)\right)$.
%
{}When the mobility is large enough (e.g. $v(n)=\omega\left(\sqrt{\frac{\log n}{n}}\right)$),
it plays the role of increasing the transmission radius, thus resulting
in the speedup. It can be easily verified, however, that the universal
lower bound on the spreading time is $\Theta\left(\log n\right)$,
which can only be achieved in the presence of full mobility. To summarize,
while the speedup $\Omega\left(v(n)\sqrt{n}/\log^{2}n)\right)$ can
be achieved in the regime $\sqrt{\frac{32\log n}{n}}\leq v(n)\leq\Theta(1)$,
RANDOM PUSH cannot achieve near-optimal spreading time $O\left(\text{poly}\left(\log n\right)\right)$
for single-message dissemination unless full mobility is present.

\subsubsection{Multi-Message Dissemination in Static Networks with RANDOM PUSH}

Now we turn to multi-message spreading over static networks with uncoded
random gossiping. Our analysis is developed for the regime where there
are $k$ distinct messages that satisfies $k=\omega\left(\text{polylog}\left(n\right)\right)$,
which subsumes most multi-spreading cases of interest. For its complement
regime where $k=O(\text{polylog}\left(n\right))$, an apparent lower
bound $\Omega\left(\sqrt{n}/\text{polylog}(n)\right)$ on the spreading
time can be obtained by observing that the diameter of the underlying
graph with transmission radius $\Theta\left(\sqrt{\frac{\log n}{n}}\right)$
is at least $\Omega\left(\sqrt{n}/\text{polylog}(n)\right)$. This
immediately indicates a gap $\Omega\left(\sqrt{n}/\text{polylog}(n)\right)$
between the spreading time and the lower limit $k=O\left(\text{polylog}(n)\right)$. 

The spreading time in the regime $k=\omega\left(\text{poly}\left(\log n\right)\right)$
is formally stated in Theorem \ref{thm-Random-Push-Static}, which
implies that simple RANDOM GOSSIP is inefficient in static wireless
networks, under a message injection scenario where users start message
dissemination sequentially. The setting is as follows: $(k-1)$ of
the sources inject their messages into the network at some time prior
to the $k$-th source. At a future time when each user in the network
has at least $w=\omega\left(\text{poly}\left(\log n\right)\right)$
messages, the $k$-th message (denoted by $M^{*}$) is injected into
the network. This pattern occurs, for example, when a new message
is injected into the network much later than other messages, and hence
all other messages have been spread to a large number of users. We
will show that without mobility, the spreading time under MOBILE PUSH
in these scenarios is at least of the same order as that under RANDOM
PUSH %
\footnote{We note that this section is devoted to showing the spreading inefficiency
under two uncoded one-sided push-only protocols. It has recently been
shown in \cite{Hae2011} that a network coding approach can allow
the optimal spreading time $\Theta(k)$ to be achieved over static
wireless networks or random geometric graphs.%
}, which is a polynomial factor away from the universal lower limit
$\Theta(k)$. In fact, the individual message flooding operation of
MOBILE PUSH does not accelerate spreading since each source has only
$O(\text{poly}\left(\log n\right))$ potential neighbors to communicate.

The main objective of analyzing the above scenario is to uncover the
fact that uncoded one-sided random gossiping fails to achieve near-optimal
spreading for a large number of message injection scenarios over static
networks. This is in contrast to mobile networks, where protocols
like MOBILE PUSH with the assistance of mobility is robust to all
initial message injection patterns and can always achieve near-optimal
spreading, as will be shown later. 

\begin{theorem}\label{thm-Random-Push-Static}

Assume that a new message $M^{*}$ arrives in a static network later
than other $k-1$ messages, and suppose that $M^{*}$ is first injected
into the network from a state such that each node has received at
least $w=\omega\left(\text{poly}\left(\log n\right)\right)$ distinct
messages. Denote by $T^{*}$ the time until every user receives $M^{*}$
using RANDOM PUSH, then for any constant $\epsilon>0$ we have\begin{equation}
T^{*}>w^{1-\epsilon}\sqrt{\frac{n}{128\log n}}\end{equation}
 with probability exceeding $1-n^{-2}$.

\end{theorem}

\begin{remark}Our main goal is to characterize the spreading inefficiency
when each node has received a few messages, which becomes most
significant when each has received $\Theta\left(k\right)$ messages.
In contrast, when only a constant number of messages are available
at each user, the evolution can be fairly fast since the piece selection
has not yet become a bottleneck. Hence, we consider $w=\omega(\text{polylog}(n))$,
which captures most of the spreading-inefficient regime $\left(\omega(\text{polylog}(n))\leq w\leq\Theta(k)\right)$.
The spreading can be quite slow for various message-injection process
over static networks, but can always be completed within $O\left(k\log^{2}n\right)$
with the assistance of mobility $v(n)=\omega\left(\sqrt{\log n/k}\right)$
regardless of the message-injection process, as will be shown in Theorem
\ref{thm:DiscreteMP}.

%
{}

\end{remark}

Theorem \ref{thm-Random-Push-Static} implies that if $M^{*}$ is
injected into the network when each user contains $\omega\left(\frac{k^{1+2\epsilon}}{\sqrt{n}\text{poly}\left(\log n\right)}\right)$
messages for any $\epsilon>0$, then RANDOM PUSH is unable to approach
the fastest possible spreading time $\Theta\left(k\right)$. In particular,
if the message is first transmitted when each user contains $\Omega\left(k/\text{poly}(\log n)\right)$
messages, then at least $\Omega\left(k^{1-\epsilon}\sqrt{n}/\text{poly}(\log n)\right)$
time slots are required to complete spreading. Since $\epsilon$ can
be arbitrarily small, there may exist a gap as large as $\Omega\left(\frac{\sqrt{n}}{\text{poly}(\log n)}\right)$
between the lower limit $\Theta(k)$ and the spreading time using
RANDOM PUSH. The reason is that as each user receives many distinct
messages, a bottleneck of spreading rate arises due to the low piece-selection
probability assigned for each message. A number of transmissions are
wasted due to the blindness of the one-sided message selection, which
results in capacity loss and hence largely constrain how efficient
information can flow across the network. The copies of each message
tend to cluster around the source -- the density of the copies decays
rapidly with the distance to the source. Such inefficiency becomes
more severe as the evolution proceeds, because each user will assign
an increasingly smaller piece-selection probability for each message.

\subsubsection{Multi-Message Dissemination in Mobile Networks with MOBILE PUSH}

Although the near-optimal spreading time $O\left(\text{polylog}(n)\right)$
for single message dissemination can only be achieved when there is
near-full mobility $v(n)=\Omega\left(1/\text{polylog}(n)\right)$,
a limited degree of velocity turns out to be remarkably helpful in
the multi-message case as stated in the following theorem.

\begin{theorem}\label{thm:DiscreteMP}Assume that the velocity obeys:
$v(n)=\omega\left(\sqrt{\frac{\log n}{k}}\right)$, where the number
of distinct messages obeys $k=\omega\left(\text{poly}\left(\log n\right)\right)$.
MOBILE PUSH message selection strategy is employed along with unicast
transmission strategy. Let $T_{\mathrm{mp}}^{\mathrm{uc}}(i)$ be
the time taken for all users to receive message $M_{i}$ after $M_{i}$
is first injected into the network, then with probability at least
$1-n^{-2}$ we have\begin{equation}
\forall i,\quad T_{\mathrm{mp}}^{\mathrm{uc}}(i)=O\left(k\log^{2}n\right).\end{equation}

\end{theorem}

Since each node can receive at most one message in each time slot,
the spreading time is lower bounded by $\Theta\left(k\right)$ for
any graph. Thus, our strategy with limited velocity spreads the information
essentially as fast as possible. Intuitively, this is due to the fact
that the velocity (even with restricted magnitude) helps \textit{uniformize}
the locations of all copies of each message, which significantly increases
the conductance of the underlying graph in each slot. Although the
velocity is significantly smaller than full mobility (which simply
results in a complete graph), the relatively low mixing time helps
to approximately achieve the same objective of uniformization. On
the other hand, the low spreading rates in static networks arise from
the fact that the copies of each message tend to cluster around the
source at any time instant, which decreases the number of flows going
towards new users without this message.

\begin{remark}Note that there is a $O\left(\log^{2}n\right)$ gap
between this spreading time and the lower limit $\Theta(k)$. We conjecture
that $\Theta\left(k\log n\right)$ is the exact order of the spreading
time, where the logarithmic gap arises from the blindness of peer
and piece selection. A gap of $\Theta\left(\log n\right)$ was shown
to be indispensable for complete graphs when one-sided random push
is used \cite{DebMedCho06}. Since the mobility model simply mimics
the evolution in complete graphs, a logarithmic gap appears to be
unavoidable when using our algorithm. Nevertheless, we conjecture
that with a finer tuning of the concentration of measure techniques,
the current gap $O\left(\log^{2}n\right)$ can be narrowed to $\Theta\left(\log n\right)$.
See Remark \ref{remark-Gap}.

%
{}

\end{remark}

{}

\section{Proofs and Discussions of Main Results\label{sec:Discrete-Jump-Model}}

The proofs of Theorem \ref{thm:DiscreteSP}-\ref{thm:DiscreteMP}
are provided in this section. Before continuing, we would like to
state some preliminaries regarding the mixing time of a random walk
on a 2-dimensional grid, some related concentration results, and a
formal definition of conductance.

\subsection{Preliminaries \label{sub:Preliminaries}}

\subsubsection{Mixing Time}

Define the probability of a typical node moving to subsquare $A_{i}$
at time $t$ as $\pi_{i}(t)$ starting from any subsquare, and denote
by $\pi_{i}$ the steady-state probability of a node residing in subsquare
$A_{i}$. Define the \textit{mixing time} of our random walk mobility
model as $T_{\text{mix}}\left(\epsilon\right):=\min\left\{ t:\left|\pi_{i}(t)-\pi_{i}\right|\leq\epsilon,\forall i\right\} $,
which characterizes the time until the underlying Markov chain is
close to its stationary distribution. It is well known that the mixing
time of a random walk on a grid satisfies (e.g. see \cite[Corollary 2.3]{ChenGunes07}
and \cite[Appendix C]{ChenGunes07}): \begin{equation}
T_{\text{mix}}\left(\epsilon\right)\leq\hat{c}m\left(\log\frac{1}{\epsilon}+\log n\right)\end{equation}
 for some constant $\hat{c}$. We take $\epsilon=n^{-10}$ throughout
this paper, so $T_{\text{mix}}\left(\epsilon\right)\leq c_{0}m\log n$
holds with $c_{0}=10\hat{c}$. After $c_{0}m\log n$ amount of time
slots, all the nodes will reside in any subsquare \textit{almost uniformly
likely}. In fact, $n^{-10}$ is very conservative and a much larger
$\epsilon$ suffices for our purpose, but this gives us a good sense
of the sharpness of the mixing time order. See \cite[Section 6]{RandomGraphDynamics}
for detailed characterization of the mixing time of random walks on
graphs.

\subsubsection{Concentration Results}

The following concentration result is also useful for our analysis.

\begin{lem}\label{lemmaConcentration-3} Assume that $b$ nodes are
thrown independently into $m$ subsquares. Suppose for any subsquare
$A_{i}$, the probability $q_{A_{i}}$ of each node being thrown to
$A_{i}$ is bounded as\begin{equation}
\left|q_{A_{i}}-\frac{1}{m}\right|\leq\frac{1}{3m}.\end{equation}
 Then for any constant $\epsilon$, the number of nodes $N_{A_{i}}(t)$
falling in any subsquare $A_{i}$$(1\leq i\leq m<n)$ at any time
$t\in\left[1,n^{2}\right]$ satisfies

a) if $b=\Theta\left(m\log n\right)$ and $b>32m\log n$, then\[
\mathbb{P}\left(\forall(i,t):\frac{b}{6m}\leq N_{A(i)}(t)\leq\frac{7b}{3m}\right)\geq1-\frac{2}{n^{3}};\]

b) if $b=\omega\left(m\log n\right)$, then\[
\mathbb{P}\left(\forall(i,t):\frac{\left(\frac{2}{3}-\epsilon\right)b}{m}\leq N_{A(i)}(t)\leq\frac{\left(\frac{4}{3}+\epsilon\right)b}{m}\right)\geq1-\frac{2}{n^{3}}.\]

\end{lem}

\begin{IEEEproof}See Appendix \ref{sub:Proof-of-Lemma-Concentration}.\end{IEEEproof}

This implies that the number of nodes residing in each subsquare at
each time of interest will be reasonably close to the true mean. This
concentration result follows from standard Chernoff bounds \cite[Appendix A]{Alon2008},
and forms the basis for our analysis.

\subsubsection{Conductance}

Conductance is an isoperimetric measure that characterizes the expansion
property of the underlying graph. Consider an irreducible reversible
transition matrix $P$ with its states represented by $V\text{ }\left(\left|V\right|=n\right)$.
Assume that the stationary distribution is uniform over all states.
In spectral graph theory, the conductance associated with $P$ is
\cite{RandomGraphDynamics} \begin{equation}
\Phi\left(P\right)=\inf_{B\subset V,\left|B\right|\leq\frac{n}{2}}\frac{\sum_{i\in B,j\in B^{c}}P_{ij}}{\left|B\right|},\end{equation}
which characterizes how easy the probability flow can cross from one
subset of nodes to its complement. If the transition matrix $P$ is
chosen such that\begin{equation}
P_{ij}=\begin{cases}
\frac{1}{d_{i}}, & \quad\text{if }j\in\text{neighbor}(i),\\
0, & \quad\text{else},\end{cases}\label{eq:TransitionMatrix}\end{equation}
where $d_{i}$ denotes the degree of vertex $i$, then the conductance
associated with random geometric graph with radius $r(n)$ obeys $\Phi(P)=\Theta\left(r(n)\right)$
\cite{ChenGunes07}, where $r(n)$ is the transmission radius.

\subsection{Single-message Dissemination in Mobile Networks}

We only briefly sketch the proof for Theorem \ref{thm:DiscreteSP}
in this paper, since the approach is somewhat standard (see \cite{GossipTutorialShah}).
Lemma \ref{lemmaConcentration-3} implies that with high probability,
the number of nodes residing in each subsquare will exhibit sharp
concentration around the mean $n/m$ once $n>32m\log n$. For each
message $M_{i}$, denote by $N_{i}(t)$ the number of users containing
$M_{i}$ at time $t$. The spreading process is divided into $2$
phases: $1\leq N_{i}(t)\leq n/2$ and $n/2<N_{i}(t)\leq n$. 

As an aside, if we denote by $p_{lj}$ the probability that $l$ successfully
transmit to $j$ in the next time slot conditional on the event that there are $\Theta(n/m)$ users residing in each subsquare, then if $m<\frac{n}{32\log n}$,
one has \[
p_{lj}=\begin{cases}
\Theta\left(\frac{m}{n}\right),\quad & \text{if }l\text{ and }j\text{ }\text{can move to the same}\\
 & \text{subsquare in the next time slot},\\
0. & \text{else}.\end{cases}\]
Concentration results imply that for any given time $t$ and any user
$l$, there are $\Theta\left(\frac{n}{m}\right)$ users that can lie
within the same subsquare as $l$ with high probability. On the other
hand, for a geometric random graph with $r(n)=\sqrt{1/m}=v(n)$, the
transition matrix defined in (\ref{eq:TransitionMatrix}) satisfies
$P_{lj}=\Theta\left(\frac{m}{n}\right)$ for all $j$ inside the transmission
range of $l$ (where there are with high probability $\Theta\left(\frac{n}{m}\right)$
users inside the transmission range). Therefore, if we define the
\emph{conductance related to this mobility model} as $\Phi\left(n\right)=\inf_{B\subset V,\left|B\right|\leq\frac{n}{2}}\frac{\sum_{i\in B,j\in B^{c}}p_{lj}}{\left|B\right|}$,
then this is order-wise equivalent to the conductance of the geometric
random graph $ $with $r(n)=v(n)$, and hence $\Phi(n)=\Theta(r(n))=\Theta(v(n))$.

\subsubsection{Phase 1}

Look at the beginning of each slot, all senders containing $M_{i}$
may transmit it to any nodes in the $9$ subsquares equally likely
with constant probability by the end of this slot. Using the same
argument as \cite{GossipTutorialShah}, one can see that the expected
increment of $N_{i}(t)$ by the end of this slot can be lower bounded
by the number of nodes $N_{i}(t)$ times the conductance related to
the mobility model $\Phi(n)=\Theta\left(v(n)\right)$ defined above.
We can thus conclude that before $N(t)=n/2$, \[
\mathbb{E}\left(N_{i}(t+1)-N_{i}(t)\mid N_{i}(t)\right)\geq b_{1}N_{i}(t)\Phi\left(n\right)=\tilde{b}_{1}N_{i}(t)v(n)\]
 holds for some constant $b_{1}$ and $\tilde{b}_{1}$. Following
the same martingale-based proof technique used for single-message
dissemination in \cite[Theorem 3.1]{GossipTutorialShah}, we can prove
that for any $\epsilon>0$, the time $T_{i1}(\epsilon)$ by which
$N_{i}(t)\geq n/2$ holds with probability at least $1-\epsilon$
can be bounded by \begin{equation}
T_{i1}(\epsilon)=O\left(\frac{\log n+\log\epsilon^{-1}}{\Phi(n)}\right)=O\left(\frac{\log n+\log\epsilon^{-1}}{v(n)}\right).\end{equation}
 Take $\epsilon=n^{-3}$, then $T_{i1}(\epsilon)$ is bounded by $O\left(\frac{\log n}{v(n)}\right)$
with probability at least $1-n^{-3}$.

\subsubsection{Phase 2}

This phase starts at $T_{i1}$ and ends when $N_{i}(t)=n$. Since
the roles of $j$ and $l$ are symmetric, the probability of $j$
having $l$ as the nearest neighbor is equal to the probability of
$l$ having $j$ as the nearest neighbor. This further yields $p_{lj}=p_{jl}$
by observing that the transmission success probability is the same
for each designated pair. Therefore, we can see: \begin{align}
 & \mathbb{E}\left(N_{i}(t+1)-N_{i}(t)\mid N_{i}(t)\right)\nonumber \\
\geq & \enskip b_{2}\sum_{l\in\mathcal{N}_{i}(t),j\notin\mathcal{N}_{i}(t)}p_{lj}\label{eq:Phase2SinglePieceStayMove}\\
= & \enskip b_{2}\left(n-N_{i}(t)\right)\frac{\sum_{j\notin\mathcal{N}_{i}(t),l\in\mathcal{N}_{i}(t)}p_{jl}}{n-N_{i}(t)}\nonumber \\
\geq & \enskip b_{2}\left(n-N_{i}(t)\right)\Phi(n).\end{align}

Denote by $T_{i2}$ the duration of Phase 2. We can follow the same
machinery in \cite{MoskSha08} to see:\begin{equation}
T_{i2}=O\left(\frac{\log n}{\Phi(n)}\right)=O\left(\frac{\log n}{v(n)}\right)\end{equation}
 with probability exceeding $1-n^{-3}$.

By combining the duration of Phase 1 and Phase 2 and applying the
union bound over all distinct messages, we can see that $T_{\text{sp}}^{\text{uc}}(i)=O\left(\frac{\log n}{v(n)}\right)$
holds for all distinct messages with high probability. When $v(n)>\sqrt{\frac{32\log n}{n}}$,
at any time instance each node can only transmit a message to nodes
at a distance at most $O\left(v(n)\right)$, and hence it will take
at least $\Omega\left(1/v(n)\right)$ time instances for $M_{i}$
to be relayed to the node farthest from $i$ at time $0$, which is
a universal lower bound on the spreading time. Therefore, $T_{\text{sp}}^{\text{uc}}(i)$
is only a logarithmic factor away from the fundamental lower bound
$\Omega\left(\frac{1}{v(n)}\right)$. It can be seen that the bottleneck
of this upper bound lies in the conductance of the underlying random
network. When $v(n)=\omega\left(\sqrt{\frac{\log n}{n}}\right)$,
mobility accelerates spreading by increasing the conductance. The
mixing time duration is much larger than the spreading time, which
implies that the copies of each message is still spatially constrained
in a fixed region (typically clustering around the source) without
being spread out over the entire square. We note that with full mobility,
the spreading time in single message dissemination case achieves the
universal lower bound $\Theta(\log n)$, which is much smaller than
that with a limited degree of velocity.

\subsection{Multi-message Spreading in Static Networks with RANDOM PUSH}

The proof idea of Theorem \ref{thm-Random-Push-Static} is sketched
in this subsection.

\subsubsection{The Lower Bound on the Spreading Time}

\begin{figure}[htbp]
\begin{centering}
\textsf{\includegraphics[scale=0.4]{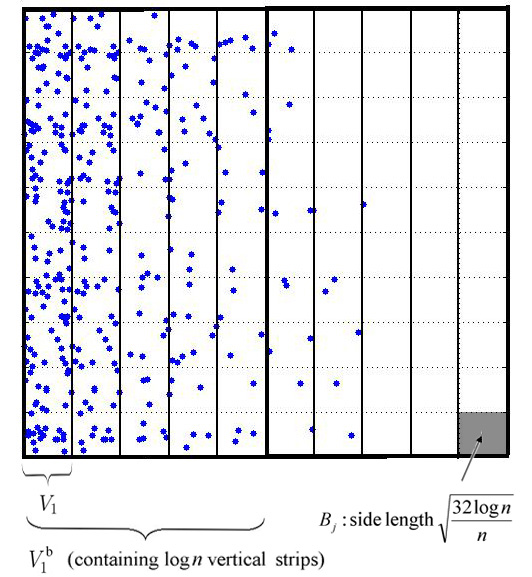}} 
\par\end{centering}

\caption{\label{fig:DecayVert}The plot illustrates that the number $N_{V_{i}}(t_{X})$
of nodes containing $M_{i}$ in vertical strip $V_{i}$ by time $t_{X}$
is decaying rapidly with geometric rate. }
\end{figure}

To begin our analysis, we partition the entire unit square as follows: 
\begin{itemize}
\item The unit square is divided into a set of nonoverlapping \textbf{\textit{tiles}}
$\left\{ B_{j}\right\} $ each of side length $\sqrt{32\log n/n}$
as illustrated in Fig. \ref{fig:DecayVert} (Note that this is a different
partition from subsquares $\left\{ A_{j}\right\} $ resulting from
the mobility model). 
\item The above partition also allows us to slice the network area into
\textbf{\textit{vertical strips}} each of width $\sqrt{32\log n/n}$
and length $1$. Label the vertical strips as $\left\{ V_{l}\right\} \left(1\leq l\leq\sqrt{n/\left(32\log n\right)}\right)$
in increasing order from left to right, and denote by $N_{V_{l}}(t)$
and $\mathcal{N}_{V_{l}}(t)$ the number and the set of nodes in $V_{l}$
that contains $M^{*}$ by time $t$. 
\item The vertical strips are further grouped into \textbf{\textit{vertical
blocks}} $\left\{ V_{j}^{\text{b}}\right\} $ each containing $\log n$
strips, i.e. $V_{j}^{\text{b}}=\left\{ V_{l}:(j-1)\log n+1\leq l\leq j\log n\right\} $. 
\end{itemize}
\begin{remark}Since each tile has an area of $32\log n/n$, concentration
results (Lemma \ref{lemmaConcentration-3}) imply that there are $\Theta\left(\log n\right)$
nodes residing in each tile with high probability. Since each sender
only attempts to transmit to its nearest receiver, then with high
probability the communication process occurs only to nodes within
the same tile or in adjacent tiles. \end{remark}

Without loss of generality, we assume that the source of $M^{*}$
resides in the \textit{leftmost} vertical strip $V_{1}$. We aim at
counting the time taken for $M^{*}$ to cross each vertical block
horizontally. In order to decouple the counting for different vertical
blocks, we construct a new spreading process $\mathcal{G}^{*}$ as
follows.

\vspace{6pt}

\framebox{%
\begin{minipage}[t]{3.2in}%
Spreading in Process $\mathcal{G}^{*}$: 
\begin{enumerate}
\item At $t=0$, distribute $M^{*}$ to all nodes residing in vertical strip
$V_{1}$. 
\item Each node adopts RANDOM PUSH as the message selection strategy. 
\item Define $T_{l}^{\text{b}}=\min\left\{ t:N_{V_{l}^{\text{b}}}(t)>0\right\} $
as the first time that $M^{*}$ reaches vertical block $V_{l}^{\text{b}}$.
For all $l\geq2$, distribute $M^{*}$ to all nodes residing in either
$V_{l-1}^{\text{b}}$ or the leftmost strip of $V_{l}^{\text{b}}$
at time $t=T_{l}^{\text{b}}$ . 
\end{enumerate}
\end{minipage}}

\vspace{6pt}
 It can be verified using a coupling approach that $\mathcal{G}^{*}$
evolves stochastically faster than the true process. By enforcing
mandatory dissemination at every $t=T_{l}^{\text{b}}$, we enable
separate counting for spreading time in different blocks -- the spreading
in $V_{l+1}^{\text{b}}$ after $T_{l+1}^{\text{b}}$ is independent
of what has happened in $V_{l}^{\text{b}}$. Roughly speaking, since
there are $\sqrt{\frac{n}{32\log^{3}n}}$ blocks, the spreading time
over the entire region is $\Theta\left(\sqrt{\frac{n}{\text{poly}\left(\log n\right)}}\right)$
times the spreading time over a typical block.

We perform a single-block analysis in the following lemma, and characterize
the rate of propagation across different strips over a typical block
in $\mathcal{G}^{*}$. By Property 3) of $\mathcal{G}^{*}$, the time
taken to cross each typical block is equivalent to the crossing time
in $V_{1}^{\text{b}}$. Specifically, we demonstrate that the time
taken for a message to cross a single block is at least $\Omega\left(w^{1-\epsilon}\right)$
for any positive $\epsilon$. Since the crossing time for each block
in $\mathcal{G}^{*}$ is statistically equivalent, this single-block
analysis further allows us to lower bound the crossing time for the
entire region. 

{}

%
{}

\begin{lem}\label{lemma-cross-strip}Consider the spreading of $M^{*}$
over $V_{1}^{\text{b}}$ in the original process $\mathcal{G}$. Suppose
each node contains at least $w=\omega\left(\text{poly}\left(\log n\right)\right)$
messages initially. Define $t_{X}:=w^{1-\epsilon}$, and define $l^{*}=\min\left\{ l:N_{V_{l}}(t_{X})=O\left(w^{\epsilon}\log n\right)\right\} $.
Then with probability at least $1-3n^{-3}$, we have

(a) $l^{*}\leq\frac{1}{2}\log n;$

(b) $\forall s$$\text{ }\left(1\leq s<l^{*}\right)$, there exists
a constant $c_{31}$ such that \begin{align}
 & N_{V_{s}}(t_{X})\leq\left(\frac{\log n}{w^{\epsilon}}\right)^{s-1}\left(c_{31}\sqrt{n}\log n\right);\label{eq:cross-strip-decay}\end{align}

(c) $N_{V_{\frac{1}{2}\log n}}(t_{X})\leq\log^{2}n$.

\end{lem}

\begin{IEEEproof}[{\bf Sketch of Proof of Lemma \ref{lemma-cross-strip}}]The
proof makes use of the fixed-point type of argument. The detailed
derivation is deferred to Appendix \ref{sec:Proof-of-Lemma-Cross-Strip}.

%
{}\end{IEEEproof}

The key observation from the above lemma is that the number of nodes
in $V_{s}$ containing $M^{*}$ is decaying rapidly as $s$ increases,
which is illustrated in Fig. \ref{fig:DecayVert}. We also observe
that $N_{V_{l}}(t_{X})$ decreases to $O\left(\log^{2}n\right)$ before
$V_{\frac{1}{2}\log n}$.

While Lemma \ref{lemma-cross-strip} determines the number of copies
of $M^{*}$ inside $V_{1}\sim V_{\frac{1}{2}\log n}$ by time $t_{X}$,
it does not indicate whether $M^{*}$ has crossed the block $V_{1}^{\text{b}}$
or not by $t_{X}$. It order to characterize the crossing time, we
still need to examine the evolution in strips $V_{\frac{1}{2}\log n+1}\sim V_{\log n}$.
Since communication occurs only between adjacent strips or within
the same strip, all copies lying to the right of $V_{\frac{1}{2}\log n}$
must be relayed via a path that starts from $V_{1}$ and passes through
$V_{\frac{1}{2}\log n}$. That said, all copies in $V_{\frac{1}{2}\log n+1}\sim V_{\log n}$
by time $t_{X}$ must have been forwarded (possibly in a multi-hop
manner) via some nodes having received $M^{*}$ by $t_{X}$. If we
denote by $\mathcal{N}_{V_{\log n/2}}^{*}\left(t_{X}\right)$ the
set of nodes in $V_{\log n/2}$ having received $M^{*}$ by $t_{X}$
in $\mathcal{G}^{*}$, then we can construct a process $\overline{\mathcal{G}}$
in which all nodes in $\mathcal{N}_{V_{\log n/2}}^{*}\left(t_{X}\right)$
receive $M^{*}$ from the very beginning ($t=0$), and hence the evolution
in $\overline{\mathcal{G}}$ can be stochastically faster than $\mathcal{G}^{*}$
by time $t_{X}$.

\vspace{6pt}

\framebox{%
\begin{minipage}[t]{3.2in}%
Spreading in Process $\overline{\mathcal{G}}$: 
\begin{enumerate}
\item Initialize (a): at $t=0$, for all $v\in\mathcal{N}_{V_{\log n/2}}^{*}(t_{X})$,
distribute $M^{*}$ to all nodes residing \textit{in the same tile}
as $v$. 
\item Initialize (b): at $t=0$, if $v_{1}$ and $v_{2}$ are two nodes
in $\mathcal{N}_{V_{\log n/2}}^{*}(t_{X})$ such that $v_{1}$ and
$v_{2}$ are less than $\log n$ tiles away from each other, then
distribute $M^{*}$ to all nodes in all tiles between $v_{1}$ and
$v_{2}$ in $V_{\log n/2}$. After this step, tiles that contain $M^{*}$
forms a set of \textit{nonoverlapping} substrips. 
\item By time $t_{X}=w^{1-\epsilon}$, the evolution to the left of $V_{\log n/2}$
occurs exactly the same as in $\mathcal{G}^{*}$. 
\item At the first time slot in which any node in the above substrips selects
$M^{*}$ for transmission, distribute $M^{*}$ to all nodes in all
tiles adjacent to any of these substrips. In other words, we expand
all substrips outwards by one tile. 
\item Repeat from 4) but consider the new set of substrips after expansion. 
\end{enumerate}
\end{minipage}}

\vspace{6pt}

By our construction of $\overline{\mathcal{G}}$, the evolution to
the left of $V_{\frac{1}{2}\log n}$ stays completely the same as
that in $\mathcal{G}^{*}$, and hence there is no \emph{successful}
transmission of $M^{*}$ between nodes in $V_{1}\sim V_{\frac{1}{2}\log n-1}$
and those in $V_{\log n/2}$ but not contained in $\mathcal{N}_{V_{\log n/2}}^{*}(t_{X})$.
Therefore, in our new process $\overline{\mathcal{G}}$, the evolution
to the left of $V_{\frac{1}{2}\log n}$ by time $t_{X}$ is decoupled
with that to the right of $V_{\frac{1}{2}\log n}$ by time $t_{X}$. 

Our objective is to examine how likely $T_{2}^{*}=\min\left\{ t:M^{*}\text{ reaches }V_{\log n}\text{ in }\overline{\mathcal{G}}\right\} $
is smaller than $t_{X}$. It can be observed that any two substrips
would never merge before $T_{2}^{*}$ since they are initially spaced
at least $\log n$ tiles from each other. This allows us to treat
them separately. Specifically, the following lemma provides a lower
bound on $T_{2}^{*}$ by studying the process $\overline{\mathcal{G}}$.

\begin{lem}\label{Lemma-Cross-Strip-Remaining}Suppose $t_{X}=w^{1-\epsilon}$
and each node contains at least $w$ distinct messages since $t=0$.
Then we have\begin{equation}
\mathbb{P}\left(T_{2}^{*}\leq t_{X}\right)\leq\frac{4}{n^{3}}.\end{equation}

\end{lem}

\begin{IEEEproof}See Appendix \ref{sec:Proof-of-Lemma-Cross-Strip-Remaining}.
\end{IEEEproof}

This lemma indicates that $M^{*}$ is unable to cross $V_{1}^{\text{b}}$
by time $t_{X}=w^{1-\epsilon}$ in $\overline{\mathcal{G}}$. Since
$\overline{\mathcal{G}}$ is stochastically faster than the original
process, the time taken for $M^{*}$ to cross a vertical block in
the original process exceeds $t_{X}$ with high probability. In other
words, the number of nodes having received $M^{*}$ by $t_{X}$ vanishes
within no more than $O\left(\log n\right)$ further strips.

Since there are $\Theta\left(\sqrt{n/\text{poly}(\log n)}\right)$
vertical blocks in total, and crossing each block takes at least $\Omega\left(w^{1-\epsilon}\right)$
time slots, the time taken for $M^{*}$ to cross all blocks can thus
be bounded below as \begin{equation}
T^{*}=\mbox{\ensuremath{\Omega}}\left(w^{1-\epsilon}\sqrt{\frac{n}{\text{poly}\left(\log n\right)}}\right)\end{equation}
 with high probability.

\subsubsection{Discussion}

Theorem \ref{thm-Random-Push-Static} implies that if a message $M^{*}$
is injected into the network when each user contains $\Omega\left(k/\text{poly}(\log n)\right)$
messages, the spreading time for $M^{*}$ is $\mbox{\ensuremath{\Omega}}\left(k^{1-\epsilon}\sqrt{n/\text{poly}(\log n)}\right)$
for arbitrarily small $\epsilon$. That said, there exists a gap as
large as $\Omega\left(\sqrt{n/\text{poly}(\log n)}\right)$ from optimality.
The tightness of this lower bound can be verified by deriving an \textit{upper
bound} using the conductance-based approach as follows.

We observe that the message selection probability for $M^{*}$ is
always lower bounded by $1/k$. Hence, we can couple a new process
adopting a different message-selection strategy such that a transmitter
containing $M^{*}$ selects it for transmission with \textit{state-independent}
probability $1/k$ at each time. It can be verified that this process
evolves stochastically slower than the original one. The conductance
associated with the new evolution for $M^{*}$ is $\Phi(n)=\frac{1}{k}\Theta\left(r(n)\right)=O\left(\frac{1}{k}\sqrt{\frac{\log n}{n}}\right)$.
Applying similar analysis as in \cite{MoskSha08} yields\begin{equation}
T_{i}=O\left(\frac{\text{poly}\left(\log n\right)}{\Phi(n)}\right)=O\left(k\sqrt{n}\text{poly}(\log n)\right)\end{equation}
 with probability exceeding $1-n^{-2}$, which is only a poly-logarithmic
gap from the lower bound we derived.

\begin{figure}[htbp]
\begin{centering}
\textsf{\includegraphics[scale=0.43]{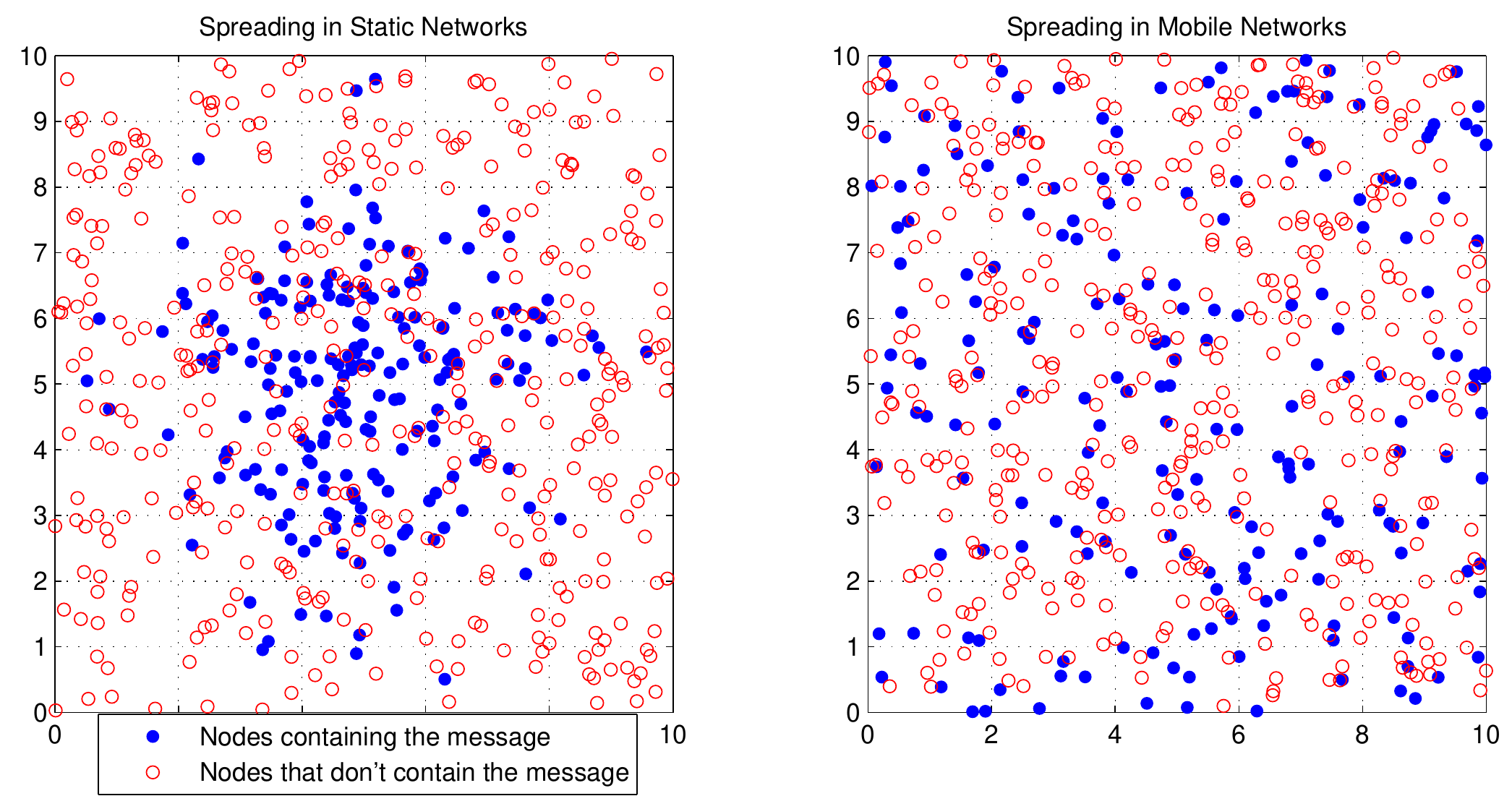}} 
\par\end{centering}

\caption{\label{fig:SpreadOut}The left plot illustrates the clustering phenomena
of $\mathcal{N}_{i}(t)$ in the evolution in a static network. However,
even restricted mobility may allow these nodes to spread out within
the mixing time duration as illustrated in the right plot. }
\end{figure}

The tightness of this upper bound implies that the propagation bottleneck
is captured by the conductance-based measure -- the copies of each
message tend to cluster around the source at any time instead of spreading
out (see Fig. \ref{fig:SpreadOut}). That said, only the nodes lying
around the boundary are likely to forward the message to new users.
Capacity loss occurs to the users inside the cluster since many transmissions
occur to receivers who have already received the message and are thus
wasted. This graph expansion bottleneck can be overcome with the assistance
of mobility.

\subsection{Multi-message Spreading in Mobile Networks with MOBILE PUSH}

The proof of Theorem \ref{thm:DiscreteMP} is sketched in this subsection.
We divide the entire evolution process into 3 phases. The duration
of Phase 1 is chosen to allow each message to be forwarded to a sufficiently
large number of users. After this initial phase (which acts to {}``seed''
the network with a sufficient number of all the messages), random
gossiping ensures the spread of all messages to all nodes.

\subsubsection{Phase 1}

This phase accounts for the first $c_{6}\left(c_{0}m\log n+\frac{m}{c_{\text{h}}\log n}\right)\log^{2}n=\Theta\left(m\log^{3}n\right)$
time slots, where $c_{0}$, $c_{6}$ and $c_{\text{h}}$ are constants
independent of $m$ and $n$. At the end of this phase, each message
will be contained in at least $32m\log n=\Theta\left(m\log n\right)$
nodes. The time intended for this phase largely exceeds the mixing
time of the random walk mobility model, which enables these copies
to {}``uniformly'' spread out over space.

We are interested in counting how many nodes will contain a particular
message $M_{i}$ by the end of Phase 1. Instead of counting all potential
multi-hop relaying of $M_{i}$, we only look at the set of nodes that
receive $M_{i}$ \textit{directly} from source $i$ in \textit{odd}
slots. This approach provides a crude lower bound on $N_{i}(t)$ at
the end of Phase 1, but it suffices for our purpose.

Consider the following scenario: at time $t_{1}$, node $i$ attempts
to transmit its message $M_{i}$ to receiver $j$. Denote by $Z_{i}(t)$
$(1\leq i\leq n)$ the subsquare position of node $i$, and define
the relative coordinate $Z_{ij}(t):=Z_{i}(t)-Z_{j}(t)$. Clearly,
$Z_{ij}(t)$ forms another two-dimensional random walk on a discrete
torus. For notional convenience, we introduce the the notation $\mathbb{P}_{0}\left(\cdot\right)\overset{\Delta}{=}\mathbb{P}\left(\cdot\mid Z_{ij}(0)=\left(0,0\right)\right)$
to denote the \emph{conditional measure} given $Z_{ij}\left(0\right)=\left(0,0\right)$.
The following lemma characterizes the hitting time of this random
walk to the boundary.

\begin{lem}\label{lemHittingTime}Consider the symmetric random walk
$Z_{ij}(t)$ defined above. Denote the set $\mathcal{A}_{\mathrm{bd}}$
of subsquares on the boundary as\[
\mathcal{A}_{\mathrm{bd}}=\left\{ A_{i}\left|A_{i}=\left(\pm\frac{\sqrt{m}}{2},j\right)\right.\text{ or }A_{i}=\left(j,\pm\frac{\sqrt{m}}{2}\right),\forall j\right\} ,\]
and define the first hitting time to the boundary as $T_{\mathrm{hit}}=\min\left\{ t:Z_{ij}(t)\in\mathcal{A}_{\mathrm{bd}}\right\} $,
then there is a constant $c_{h}$ such that\begin{equation}
\mathbb{P}_{0}\left(T_{\mathrm{hit}}<\frac{m}{c_{h}\log n}\right)\leq\frac{1}{n^{4}}.\end{equation}

\end{lem}

\begin{IEEEproof}See Appendix \ref{sec:Proof-of-Lemma-Hitting-Time}.\end{IEEEproof}

Besides, the following lemma provides an upper bound on the expected
number of time slots by time $t$ during which the walk returns to
$(0,0)$.

\begin{lem}\label{lemSingleWalk}For the random walk $Z_{ij}(t)$
defined above, there exist constants $c_{3}$ and $c_{\text{h}}$
such that for any $t<\frac{m}{c_{\text{h}}\log n}$: \begin{align}
\mathbb{E}\left(\left.\sum_{k=1}^{t}\mathds{1}\left(Z_{ij}(k)=(0,0)\right)\right|Z_{ij}(0)=(0,0)\right)\leq c_{3}\log t.\label{eq:RandomWalkReturnTimes}\end{align}
 Here, $\mathds{1}\left(\cdot\right)$ denotes the indicator function.

\end{lem}

\begin{IEEEproof}[{\bf Sketch of Proof of Lemma \ref{lemSingleWalk}}]Denote
by $\mathcal{H}_{\text{bd}}$ the event that $Z_{ij}(t)$ hits the
boundary $\mathcal{A}_{\text{bd}}$ (as defined in Lemma \ref{lemHittingTime})
before $t=m/\left(c_{\text{h}}\log n\right)$. Conditional on $Z_{ij}(0)=\left(0,0\right)$,
the probability $q_{ij}^{0}(t)$ of $Z_{ij}(t)$ returning to $(0,0)$
at time $t$ can then be bounded as\begin{equation}
q_{ij}^{0}(t)\leq\mathbb{P}_{0}\left(\mathcal{H}_{\text{bd}}\right)+\mathbb{P}_{0}\left(Z_{ij}(t)=\left(0,0\right)\wedge\overline{\mathcal{H}}_{\text{bd}}\right).\end{equation}
 Now, observe that when restricted to the set of sample paths where
$Z_{ij}(t)$ does not reach the boundary by $t$, we can couple the
sample paths of $Z_{ij}(t)$ to the sample paths of a random walk
$\tilde{Z}_{ij}(t)$ over an infinite plane before the corresponding
hitting time to the boundary. Denote by $\tilde{\mathcal{H}}_{\text{bd}}$
the event that $\tilde{Z}_{ij}(t)$ hits $\mathcal{A}_{\text{bd}}$
by $t=m/\left(c_{\text{h}}\log n\right)$ , then \begin{align*}
\mathbb{P}_{0}\left(Z_{ij}(t)=\left(0,0\right)\wedge\overline{\mathcal{H}}_{\text{bd}}\right) & =\mathbb{P}_{0}\left(\tilde{Z}_{ij}(t)=\left(0,0\right)\wedge\overline{\tilde{\mathcal{H}}}_{\text{bd}}\right)\\
 & \leq\mathbb{P}_{0}\left(\tilde{Z}_{ij}(t)=\left(0,0\right)\right).\end{align*}
 The return probability obeys $\mathbb{P}_{0}\left(\tilde{Z}_{ij}(t)=\left(0,0\right)\right)\sim t^{-1}$
for a random walk over an infinite plane \cite{FosterGood53}, and
$\mathbb{P}_{0}\left(\mathcal{H}_{\text{bd}}\right)$ will be bounded
in Lemma \ref{lemHittingTime}. Summing up all $q_{ij}^{0}(t)$ yields
(\ref{eq:RandomWalkReturnTimes}). See Appendix \ref{sec:Proof-of-Lemma-SingleWalk}
for detailed derivation.\end{IEEEproof}

In order to derive an estimate on the number of distinct nodes receiving
$M_{i}$ directly from source $i$, we need to calculate the number
of slots where $i$ fails to forward $M_{i}$ to a new user. In addition
to physical-layer outage events, some transmissions occur to users
already possessing $M_{i}$, and hence are not successful. Recall
that we are using one-sided push-only strategy, and hence we cannot
always send an innovative message. Denote by $F_{i}\left(t\right)$
the number of \textit{wasted transmissions} from $i$ to some users
already containing $M_{i}$ by time $t$. This can be estimated as
in the following lemma.

\begin{lem}\label{lemma-NumOfFailureBcRetransmissions} For $t_{0}=\frac{m}{c_{\text{h}}\log n}$,
the number of \textit{wasted transmissions} $F_{i}(t)$ defined above
obeys \begin{equation}
\mathbb{E}\left(F_{i}(t_{0})\right)\leq c_{5}\frac{m\log n}{n}t_{0}\end{equation}
 for some fixed constant $c_{5}$ with probability exceeding $1-3n^{-3}$.

\end{lem}

\begin{IEEEproof}[{\bf Sketch of Proof of Lemma \ref{lemma-NumOfFailureBcRetransmissions}}]Consider
a particular pair of nodes $i$ and $j$, where $i$ is the source
and $j$ contains $M_{i}$. A wasted transmission occurs when (a)
$i$ and $j$ meets in the same subsquare again, and (b) $i$ is designated
as a sender with $j$ being the intended receiver. The probability
of event (a) can be calculated using Lemma \ref{lemSingleWalk}. Besides,
the probability of (b) is $\Theta\left(m/n\right)$ due to sharp concentration
on $N_{A_{i}}$. See Appendix \ref{sec:Proof-of-Lemma-Number-Of-Fairlure}.
\end{IEEEproof}

The above result is helpful in estimating the \textit{expected number}
of distinct users containing $M_{i}$. However, it is not obvious
whether $F_{i}(t)$ exhibits desired sharp concentration. The difficulty
is partly due to the dependence among $\left\{ Z_{ij}(t)\right\} $
for different $t$ arising from its Markov property. Due to their
underlying relation with location of $i$, $Z_{ij_{1}}(t)$ and $Z_{ij_{2}}(t)$
are not independent either for $j_{1}\neq j_{2}$. However, this difficulty
can be circumvented by constructing different processes that exhibit
approximate mutual independence as follows.

\begin{figure}[htbp]
\begin{centering}
\textsf{\includegraphics[scale=0.8]{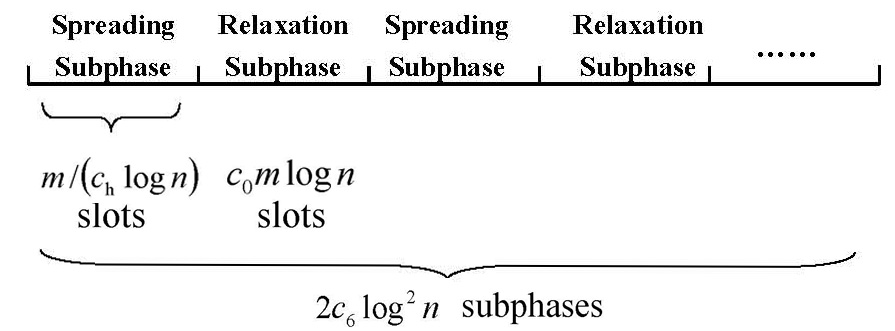}} 
\par\end{centering}

\caption{\label{fig:Phase1Subphase}Phase 1 is divided into $2c_{6}\log^{2}n$
subphases. Each odd subphase accounts for $m/\left(c_{\text{h}}\log n\right)$
slots, during which all nodes perform message spreading. Each even
subphase contains $c_{0}m\log n$ slots, during which no transmissions
occur; it allows all nodes containing a typical message to be uniformly
spread out. }
\end{figure}

The time duration $\left[1,c_{6}\left(c_{0}m\log n+m/\left(c_{\text{h}}\log n\right)\right)\log^{2}n\right]$
of Phase 1 are divided into $c_{6}\log^{2}n$ non-overlapping subphases
$P_{1,j}$ $\left(1\leq j\leq\log^{2}n\right)$ for some constant
$c_{6}$. Each odd subphase accounts for $m/\left(c_{\text{h}}\log n\right)$
time slots, whereas each even subphase contains $c_{0}m\log n$ slots.
See Fig. \ref{fig:Phase1Subphase} for an illustration. Instead of
studying the true evolution, we consider different evolutions for
each subphase. In each odd subphase, source $i$ attempts to transmit
message $M_{i}$ to its intended receiver as in the original process.
But in every even subphase, all new transmissions will be immediately
deleted. The purpose for constructing these \textit{clearance} or
\textit{relaxation} processes in even subphases is to allow for approximately
independent counting for odd subphases. The duration $c_{0}m\log n$
of each even subphase, which is larger than the typical mixing time
duration of the random walk, is sufficient to allow each user to move
to everywhere almost uniformly likely.

\begin{lem}\label{lemma-Phase1-Ni-Bound} Set $t$ to be $c_{6}\left(c_{0}m\log n+\frac{m}{c_{\text{h}}\log n}\right)\log^{2}n$,
which is the end time slot of Phase 1. The number of users containing
each message $M_{i}$ can be bounded below as

\begin{equation}
\forall i,\quad N_{i}\left(t\right)>32m\log n\end{equation}
 with probability at least $1-c_{7}n^{-2}$.

\end{lem}

\begin{IEEEproof}See Appendix \ref{sec:Proof-of-Lemma-Phase1-Ni-Bound}.\end{IEEEproof}

In fact, if $m\log^{2}n\ll n$ holds, the above lemma can be further
refined to $N_{i}\left(t\right)=\Theta\left(m\log^{2}n\right)$. This
implies that, by the end of Phase 1, each message has been flooded
to $\Omega\left(m\log n\right)$ users. They are able to \textit{cover}
all subsquares (i.e., the messages' locations are roughly uniformly
distributed over the unit square) after a further mixing time duration.

\subsubsection{Phase 2}

This phase starts from the end of Phase 1 and ends when $N_{i}(t)>n/8$
for all $i$. We use $t=0$ to denote the starting slot of Phase 2
for convenience of presentation. Instead of directly looking at the
original process, we generate a new process $\tilde{\mathcal{G}}$
which evolves slower than the original process $\mathcal{G}$. Define
$\mathcal{S}_{i}(t)$ and $\tilde{\mathcal{S}}_{i}(t)$ as the set
of messages that node $i$ contains at time $t$ in $\mathcal{G}$
and $\tilde{\mathcal{G}}$, with $S_{i}(t)$ and $\tilde{S}_{i}(t)$
denoting their cardinality, respectively. For more clear exposition,
we divide the entire phase into several time blocks each of length
$k+c_{0}\log n/v^{2}(n)$, and use $t_{B}$ to label different time
blocks. We define $\tilde{\mathcal{N}}_{i}^{B}(t_{B})$ to denote
$\tilde{\mathcal{N}}_{i}(t)$ with $t$ being the starting time of
time block $t_{B}$. $\tilde{\mathcal{G}}$ is generated from $\mathcal{G}$:
everything in these two processes remains the same (including locations,
movements, physical-layer outage events, etc.) except message selection
strategies, detailed below:

\vspace{6pt}

\framebox{%
\begin{minipage}[t]{3.2in}%
Message Selection Strategy in the Coupled Process $\tilde{\mathcal{G}}$: 
\begin{enumerate}
\item ${\bf \text{Initialize}}$: At $t=0$, for all $i$, copy the set
$\mathcal{S}_{i}(t)$ of all messages that $i$ contains to $\tilde{\mathcal{S}}_{i}(t)$.
Set $t_{B}=0$. 
\item In the next $c_{0}\log n/v^{2}(n)$ time slots, all new messages received
in this subphase are immediately deleted, i.e., no successful forwarding
occurs in this subphase regardless of the locations and physical-layer
conditions. 
\item In the next $k$ slots, for every sender $i$, each message it contains
is randomly selected with probability $1/k$ for transmission. 
\item For all $i$, if the number of nodes containing $M_{i}$ is larger
than $2\tilde{N}_{i}^{B}(t_{B})$, delete $M_{i}$ from some of these
nodes so that $\tilde{N}_{i}(t)=2\tilde{N}_{i}^{B}(t_{B})$ by the
end of this time block. 
\item Set $t_{B}\leftarrow t_{B}+1$. Repeat from (2) until $N_{i}>n/8$
for all $i$. 
\end{enumerate}
\end{minipage}}

\vspace{6pt}

Thus, each time block consists of a relaxation period and a spreading
period. The key idea is to simulate an \textit{approximately spatially-uniform
evolution}, which is summarized as follows: 
\begin{itemize}
\item After each spreading subphase, we give the process a \textit{relaxation}
period to allow each node to move almost uniformly likely to all subsquares.
This is similar to the relaxation period introduced in Phase 1. 
\item Trimming the messages alone does \textit{not} necessarily generate
a slower process, because it potentially increases the selection probability
for each message. Therefore, we force the message selection probability
to be a lower bound $1/k$, which is \textit{state-independent}. Surprisingly,
this conservative bound suffices for our purpose because it is exactly
one of the \textit{bottlenecks} for the evolution. 
\end{itemize}
The following lemma makes a formal comparison of $\mathcal{G}$ and
$\tilde{\mathcal{G}}$.

\begin{lem}\label{lemma-StochasticOrderPhase2}$\tilde{\mathcal{G}}$
evolves stochastically slower than $\mathcal{G}$, i.e.\begin{equation}
\mathbb{P}\left(T_{2}>x\right)<\mathbb{P}\left(\tilde{T}_{2}>x\right),\quad\forall x>0\end{equation}
 where $T_{2}=\min\left\{ t:N_{i}(t)>n/8,\forall i\right\} $ and
$\tilde{T}_{2}=\min\left\{ t:\tilde{N}_{i}(t)>n/8,\forall i\right\} $
are the stopping time of Phase 2 for $\mathcal{G}$ and $\tilde{\mathcal{G}}$,
respectively. \end{lem}

\begin{IEEEproof}Whenever a node $i$ sends a message $M_{k}$ to
$j$ in $\mathcal{G}$: (a) if $M_{k}\in\tilde{\mathcal{S}}_{i}$,
then $i$ selects $M_{k}$ with probability $S_{i}/k$, and a random
useless message otherwise; (b) if $M_{k}\notin\tilde{\mathcal{S}}_{i}$,
$i$ always sends a random noise message. The initial condition $\tilde{\mathcal{S}}_{i}=\mathcal{S}_{i}$
guarantees that $\tilde{\mathcal{S}}_{i}\subseteq\mathcal{S}_{i}$
always holds with this coupling method. Hence, the claimed stochastic
order holds. \end{IEEEproof}

\begin{lem}\label{lemmaPhase2}Denote by $\tilde{T}_{2}^{B}:=\min\left\{ t_{B}:\tilde{N}_{i}^{B}(t_{B})>n/8,\forall i\right\} $
the stopping time block of Phase 2 in $\tilde{\mathcal{G}}$. Then
there exists a constant $c_{14}$ independent of $n$ such that\[
\mathbb{P}\left(\tilde{T}_{2}^{B}\leq4\log_{c_{14}}n\right)\leq1-n^{-2}.\]
 \end{lem}

\begin{IEEEproof}[{\bf Sketch of Proof of Lemma \ref{lemmaPhase2}}]We
first look at a particular message $M_{i}$, and use union bound later
after we derive the concentration results on the stopping time associated
with this message. We observe the following facts: after a mixing
time duration, the number of users $N_{i,A_{k}}(t)$ containing $M_{i}$
at each subsquare $A_{k}$ is approximately \textit{uniform}. Since
$N_{i}^{B}(t_{B})$ is the lower bound on the number of copies of
$M_{i}$ across this time block, concentration results suggest that
$N_{i,A_{k}}(t)=\Omega\left(N_{i}^{B}(t_{B})/m\right)$. Observing
from the mobility model that the position of any node inside a subsquare
is \textit{i.i.d.} chosen, we can derive\begin{equation}
\mathbb{E}\left(\tilde{N}_{i}^{B}(t_{B}+1)-\tilde{N}_{i}^{B}(t_{B})\mid\tilde{\mathcal{N}}_{i}^{B}(t_{B})\right)\geq\frac{\tilde{c}_{9}}{2}\tilde{N}_{i}^{B}(t_{B})\end{equation}
 for some constant $\tilde{c}_{9}$. A standard \textit{martingale}
argument then yields an upper bound on the stopping time. See Appendix
\ref{sec:Proof-of-Lemma-Phase2} for detailed derivation. \end{IEEEproof}

This lemma implies that after at most $4\log_{c_{14}}n$ time blocks,
the number of nodes containing all messages will exceed $n/8$ with
high probability. Therefore, the duration $\tilde{T}_{2}$ of Phase
2 of $\tilde{\mathcal{G}}$ satisfies $\tilde{T}_{2}=O\left(k\log n\right)$
with high probability. This gives us an upper bound on $T_{2}$ of
the original evolution $\mathcal{G}$.

\subsubsection{Phase 3}

This phase ends when $N_{i}(t)=n$ for all $i$ with $t=0$ denoting
the end of Phase 2. Assume that $N_{i,A_{j}}(0)>\frac{n}{16m}$ for
all $i$ and all $j$, otherwise we can let the process further evolve
for another mixing time duration $\Theta\left(\log n/v^{2}(n)\right)$.

\begin{lem}\label{lemmaPhase3Unicast}Denote by $T_{3}$ the duration
of Phase 3, i.e. $T_{3}=\min\left\{ t:N_{i}(t)=n\mid N_{i}(0)\geq n/8,\forall i\right\} $.
Then there exists a constant $c_{18}$ such that\begin{equation}
\mathbb{P}\left(T_{3}\leq\frac{64}{c_{18}}k\log n\right)\geq1-\frac{15}{16n^{2}}.\end{equation}
 \end{lem}

\begin{IEEEproof}[{\bf Sketch of Proof of Lemma \ref{lemmaPhase3Unicast}}]The
random push strategies are efficient near the start (exponential growth),
but the evolution will begin to slow down after Phase 2. The concentration
effect allows us to obtain a different evolution bound as\begin{align*}
 & \mbox{\text{ }}\mathbb{E}\left(N_{i}(t+1)-N_{i}(t)\left|N_{i}(t)\right.\right)\\
=\mbox{\text{ }} & \mathbb{E}\left(n-N_{i}(t)-\left(n-N_{i}(t+1)\right)\left|N_{i}(t)\right.\right)\\
\geq\mbox{\text{ }} & \frac{c_{18}}{16k}\left(n-N_{i}(t)\right).\end{align*}
 Constructing a different submartingale based on $n-N_{i}(t)$ yields
the above results. See Appendix \ref{sec:Proof-of-Lemma-Phase3-Unicast}.
\end{IEEEproof}

\subsubsection{Discussion}

Combining the stopping time in all three phases, we can see that:
the spreading time $T_{\text{mp}}^{\text{d}}=\min\left\{ t:\forall i,N_{i}(t)=n\right\} $
satisfies\[
T_{\text{mp}}^{\text{d}}\leq O\left(\frac{\log^{3}n}{v^{2}(n)}\right)+O\left(k\log n\right)+O\left(k\log n\right)=O\left(k\log^{2}n\right).\]
 It can be observed that, the mixing time bottleneck will not be critical
in multi-message dissemination. The reason is that the mixing time
in the regime $v(n)=\omega\left(\sqrt{\frac{\log n}{k}}\right)$ is
much smaller than the optimal spreading time. Hence, the nodes have
sufficient time to spread out to everywhere. The key step is how to
seed the network with a sufficiently large number of copies at the
initial stage of the spreading process, which is accomplished by the
self-promotion phase of MOBILE PUSH.

\begin{remark}\label{remark-Gap}It can be observed that the upper
bounds on spreading time within Phase 2 and Phase 3 are order-wise
tight, since a gap of $\Omega(\log n)$ exists even for complete graphs
\cite{SanHajMas07}. The upper bound for Phase 1, however, might not
be necessarily tight. We note that the $O\left(\log^{2}n\right)$
factor arises in the analysis stated in Lemma \ref{lemma-Phase1-Ni-Bound},
where we assume that each relaxation subphase is of duration $\Theta\left(m\log n\right)$
for ease of analysis. Since we consider $\Theta\left(\log^{2}n\right)$
subphases in total, we do not necessarily need $\Theta\left(m\log n\right)$
slots for each relaxation subphase in order to allow spreading of
all copies. We conjecture that with a finer tuning of the concentration
of measures and coupling techniques, it may be possible to obtain
a spreading time of $\Theta(k\log n)$. 

\end{remark}

%
{}

\section{Concluding Remarks}

In this paper, we design a simple distributed gossip-style protocol
that achieves near-optimal spreading rate for multi-message dissemination,
with the assistance of mobility. The key observation is that random
gossiping over static geometric graphs is inherently constrained by
the expansion property of the underlying graph -- capacity loss occurs
since the copies are spatially constrained instead of being spread
out. Encouragingly, this bottleneck can indeed be overcome in mobile
networks, even with fairly limited degree of velocity. In fact, the
velocity-constrained mobility assists in achieving a large expansion
property from a long-term perspective, which simulates a spatially-uniform
evolution.

\section*{Acknowledgment}

\addcontentsline{toc}{section}{Acknowledgment} The authors would
like to thank Yudong Chen and Constantine Caramanis for helpful discussions.

\appendix

\subsection{Proof of Lemma \ref{lemmaConcentration-3}\label{sub:Proof-of-Lemma-Concentration}}

Let us look at a typical time slot $t$ at subsquare $A_{i}$. We
know that $2b/3m\leq\mathbb{E}\left(N_{A(i)}(t)\right)\leq4b/3m$.
For each node $j$, define the indicator variable $X_{j,A_{i}}(t):=\mathds{1}\left\{ j\mbox{ lies in }A_{i}\text{ at }t\right\} $.
Then, $\left\{ X_{j,A_{i}}(t):1\leq j\leq b\right\} $ forms a set
of i.i.d. random variables each satisfying $\frac{2}{3m}\leq\mathbb{P}\left(X_{j,A_{i}}(t)=1\right)\leq\frac{4}{3m}$.

Define another two sets of i.i.d. Bernoulli random variables $\left\{ X_{j,A_{i}}^{\text{ub}}(t):1\leq j\leq b\right\} $
and $\left\{ X_{j,A_{i}}^{\text{lb}}(t):1\leq j\leq b\right\} $ such
that \[
\begin{cases}
\mathbb{P}\left(X_{j,A_{i}}^{\text{ub}}(t)=1\right)=\frac{4}{3m},\\
\mathbb{P}\left(X_{j,A_{i}}^{\text{lb}}(t)=1\right)=\frac{2}{3m}.\end{cases}\]
 We also define $N_{A(i)}^{\text{ub}}(t):=\underset{1\leq j\leq b}{\sum}X_{j,A_{i}}^{\text{ub}}(t)$
and $N_{A(i)}^{\text{lb}}(t):=\underset{1\leq j\leq b}{\sum}X_{j,A_{i}}^{\text{lb}}(t)$.
The following stochastic orders can be immediately observed through
simple coupling arguments \begin{equation}
N_{A(i)}^{\text{lb}}(t)\preceq N_{A(i)}(t)\preceq N_{A(i)}^{\text{ub}}(t),\end{equation}
 i.e. for any positive $d$, we have \[
\mathbb{P}\left(N_{A(i)}^{\text{lb}}(t)>d\right)\leq\mathbb{P}\left(N_{A(i)}(t)>d\right)\leq\mathbb{P}\left(N_{A(i)}^{\text{ub}}(t)>d\right).\]

Applying Chernoff bound yields \begin{align*}
 & \mathbb{P}\left(N_{A(i)}(t)\geq\frac{(1+\epsilon)4b}{3m}\right)\\
\leq & \mbox{\text{ }}\mathbb{P}\left(\sum_{j=1}^{b}X_{j,A_{i}}^{\text{ub}}(t)\geq\frac{(1+\epsilon)4b}{3m}\right)\\
\leq & \text{ }\exp\left(-\frac{2b\epsilon^{2}}{3m}\right).\end{align*}
 We can thus observe through union bound that \begin{align*}
 & \text{ }\mathbb{P}\left(\exists(i,t)\text{ such that }N_{A(i)}(t)\geq\frac{(1+\epsilon)4b}{3m}\right)\\
\leq & \text{ }mn^{2}\mathbb{P}\left(N_{A(i)}(t)\geq\frac{(1+\epsilon)4b}{3m}\right)\\
\leq & \text{ }n^{3}\exp\left(-\frac{2b\epsilon^{2}}{3m}\right).\end{align*}

Similarly, the stochastic order implies that\begin{align*}
\mathbb{P}\left(N_{A(i)}(t)\leq\frac{(1-\epsilon)2b}{3m}\right) & \leq\mathbb{P}\left(N_{A(i)}^{\text{lb}}(t)\leq\frac{(1-\epsilon)2b}{3m}\right)\\
 & \leq\exp\left(-\frac{b\epsilon^{2}}{3m}\right).\end{align*}
 Therefore, we have \begin{align*}
 & \text{ }\mathbb{P}\left(\exists(i,t)\text{ such that }N_{A(i)}(t)\leq\frac{(1-\epsilon)2b}{3m}\right)\\
\leq & \text{ }n^{3}\exp\left(-\frac{b\epsilon^{2}}{3m}\right).\end{align*}

When $b=\omega\left(m\log n\right)$, $\exp\left(-\frac{\epsilon^{2}}{3}\frac{b}{m}\right)\ll n^{-c}$
holds for any positive constant $c$; when $b=\Theta\left(m\log n\right)$
and $b>32m\log n$, taking $\epsilon=\frac{3}{4}$ completes our proof.

\subsection{Proof of Lemma \ref{lemma-cross-strip} \label{sec:Proof-of-Lemma-Cross-Strip}}

Consider first the vertical strip $V_{l}$. Obviously, the spreading
within $V_{l}$ will be influenced by nodes residing in adjacent strips
$V_{l-1}$,$V_{l+1}$, and $V_{l}$ itself. Define a set of i.i.d.
Bernoulli random variables $\left\{ X_{v,t}\right\} $ such that \begin{equation}
\mathbb{P}\left(X_{v,t}=1\right)=\frac{1}{w}.\end{equation}
 For any $v\in\mathcal{N}_{V_{l-1}}(t_{X})\bigcup\mathcal{N}_{V_{l}}(t_{X})\bigcup\mathcal{N}_{V_{l+1}}(t_{X})$,
the probability of node $v$ selecting $M^{*}$ for transmission at
time $t$ can be bounded above by $\mathbb{P}\left(X_{v,t}=1\right)$.
Simple coupling argument yields the following stochastic order\begin{align}
 & N_{V_{l}}\left(t_{X}\right)\nonumber \\
\leq\text{ } & \sum_{t=1}^{t_{X}}\sum_{v\in\overset{l+1}{\underset{i=l-1}{\bigcup}}\mathcal{N}_{V_{i}}(t_{X})}\mathds{1}\left(v\text{ selects }M^{*}\text{ for transmission at }t\right)\nonumber \\
\preceq^{\text{st}} & \sum_{t=1}^{t_{X}}\sum_{v=1}^{N_{V_{l-1}}(t_{X})+N_{V_{l}}(t_{X})+N_{V_{l+1}}(t_{X})}X_{v,t},\label{eq:stochOrderXvtNvl}\end{align}
where $\preceq^{\text{st}}$ denotes stochastic order. If $N_{V_{l-1}}(t_{X})+N_{V_{l+1}}(t_{X})>\omega\left(w^{\epsilon}\log n\right)$,
then we have\begin{align}
 & \mbox{\text{ }}\mathbb{P}\left\{ N_{V_{l}}\left(t_{X}\right)>\frac{2}{w^{\epsilon}}\left(N_{V_{l-1}}(t_{X})+N_{V_{l+1}}(t_{X})\right)\right\} \nonumber \\
\leq & \mbox{\text{ }}\mathbb{P}\left\{ \left(1-\frac{3}{2}\frac{1}{w^{\epsilon}}\right)N_{V_{l}}^{\text{x}}\left(t_{X}\right)>\right.\nonumber \\
 & \quad\quad\quad\left.\frac{3}{2}\frac{1}{w^{\epsilon}}\left(N_{V_{l-1}}(t_{X})+N_{V_{l+1}}(t_{X})\right)\right\} \\
= & \mbox{\text{ }}\mathbb{P}\left\{ N_{V_{l}}^{\text{}}\left(t_{X}\right)>\frac{3}{2}\frac{1}{w^{\epsilon}}\left(\sum_{j=l-1}^{l+1}N_{V_{j}}(t_{X})\right)\right\} \nonumber \\
\leq & \mbox{\text{ }}\mathbb{P}\left\{ \sum_{t=1}^{t_{X}}\sum_{v=1}^{N_{V_{l-1}}(t_{X})+N_{V_{l}}(t_{X})+N_{V_{l+1}}(t_{X})}X_{v,t}>\right.\nonumber \\
 & \quad\left.\frac{3}{2}\mathbb{E}\left(\sum_{t=1}^{t_{X}}\sum_{v=1}^{N_{V_{l-1}}(t_{X})+N_{V_{l}}(t_{X})+N_{V_{l+1}}(t_{X})}X_{v,t}\right)\right\} \label{eq:StochasticOrderN2N1N3Xvt}\\
\leq & \mbox{\text{ }}\frac{1}{n^{5}},\nonumber \end{align}
 where (\ref{eq:StochasticOrderN2N1N3Xvt}) follows from the stochastic
order (\ref{eq:stochOrderXvtNvl}), and the last inequality follows
from large deviation bounds and the observation that\[
\mathbb{E}\left(\sum_{t=1}^{t_{X}}\sum_{v=1}^{N_{V_{l-1}}(t_{X})+N_{V_{l}}(t_{X})+N_{V_{l+1}}(t_{X})}X_{v,t}\right)=\omega\left(\log n\right).\]

Besides, if $N_{V_{l-1}}(t_{X})+N_{V_{l+1}}(t_{X})=O\left(w^{\epsilon}\log n\right)$,
then we have\begin{align}
 & \mbox{\text{ }}\mathbb{P}\left\{ N_{V_{l}}\left(t_{X}\right)>3\log^{2}n\right\} \nonumber \\
\leq & \mbox{\text{ }}\mathbb{P}\left\{ N_{V_{l}}\left(t_{X}\right)>\frac{2}{w^{\epsilon}}\left(N_{V_{l-1}}(t_{X})+N_{V_{l+1}}(t_{X})\right)+2\log^{2}n\right\} \nonumber \\
\leq & \mbox{\text{ }}\mathbb{P}\left\{ N_{V_{l}}\left(t_{X}\right)>\frac{1}{w^{\epsilon}}\left(\sum_{j=l-1}^{l+1}N_{V_{j}}(t_{X})\right)+\log^{2}n\right\} \nonumber \\
\leq & \text{ }\mathbb{P}\left\{ \sum_{t=1}^{t_{X}}\sum_{v=1}^{N_{V_{l-1}}(t_{X})+N_{V_{l}}(t_{X})+N_{V_{l+1}}(t_{X})}Y_{v,t}>\log^{2}n\right\} ,\label{eq:LargeDevN1N2N3}\end{align}
where $Y_{v,t}\overset{\Delta}{=}X_{v,t}-1/w$. Hence, $Y_{v,t}=1-1/w$
with probability $1/w$. Let $\overline{\beta}-1:=w^{\epsilon}\log^{2}n/\left(N_{V_{1}}(t_{X})+N_{V_{2}}(t_{X})+N_{V_{3}}(t_{X})\right)=\Omega\left(\log n\right)$.
By applying \cite[Theorem A.1.12]{Alon2008}, we can show that there
exist constants $c_{21}$ and $\tilde{c}_{21}$ such that \begin{equation}
(\ref{eq:LargeDevN1N2N3})\leq\left(\frac{c_{21}}{\log n}\right)^{\log^{2}n}\leq\left(\tilde{c}_{21}e\right)^{-\log^{2}n\cdot\log\log n}\leq\frac{1}{n^{5}}.\end{equation}

Define $l^{*}:=\min\left\{ l:N_{V_{l}}(t_{X})=O\left(w^{\epsilon}\log n\right)\right\} $
and $L^{*}:=\min\left\{ l^{*},0.5\log n\right\} $. Since $N_{V_{L^{*}+1}}(t_{X})+N_{V_{L^{*}-1}}(t_{X})\geq N_{V_{L^{*}-1}}(t_{X})=\omega\left(w^{\epsilon}\log n\right)$,
we can derive with similar spirit that\begin{align*}
N_{V_{L^{*}}}(t_{X}) & \leq\frac{2}{w^{\epsilon}}\left(N_{V_{L^{*}+1}}(t_{X})+N_{V_{L^{*}-1}}(t_{X})\right)\end{align*}
 with probability at least $1-n^{-5}$. Similarly, if $N_{V_{L^{*}}}(t_{X})+N_{V_{L^{*}+2}}(t_{X})=\omega\left(w^{\epsilon}\log n\right)$,
we can derive\begin{align*}
N_{V_{L^{*}+1}}(t_{X}) & \leq\frac{2}{w^{\epsilon}}\left(N_{V_{L^{*}}}(t_{X})+N_{V_{L^{*}+2}}(t_{X})\right)\end{align*}
 with high probability. But if $N_{V_{L^{*}}}(t_{X})+N_{V_{L^{*}+2}}(t_{X})=O\left(w^{\epsilon}\log n\right)$,
it can still be shown that\begin{equation}
N_{V_{L^{*}+1}}(t_{X})\leq\log^{2}n\leq\frac{\log n}{w^{\epsilon}}N_{V_{L^{*}-2}}(t_{X})\end{equation}
 by observing that $N_{V_{L^{*}-2}}(t_{X})=\omega\left(w^{\epsilon}\log n\right)$.
Combining all these facts yields\begin{align*}
 & N_{V_{L^{*}-1}}(t_{X})+N_{V_{L^{*}+1}}(t_{X})\\
\leq & \text{ }\frac{2}{w^{\epsilon}}\left(N_{V_{L^{*}-2}}(t_{X})+N_{V_{L^{*}}}(t_{X})\right)+\\
 & \frac{\log n}{w^{\epsilon}}\left(N_{V_{L^{*}+2}}(t_{X})+N_{V_{L^{*}}}(t_{X})+N_{V_{L^{*}-2}}(t_{X})\right)\\
\leq & \text{ }\frac{3\log n}{w^{\epsilon}}\left(N_{V_{L^{*}-2}}(t_{X})+N_{V_{L^{*}}}(t_{X})+N_{V_{L^{*}+2}}(t_{X})\right)\\
\leq & \text{ }\frac{3\log n}{w^{\epsilon}}\left(N_{V_{L^{*}-2}}(t_{X})+N_{V_{L^{*}+2}}(t_{X})\right)+\\
 & \quad\frac{6\log n}{w^{2\epsilon}}\left(N_{V_{L^{*}+1}}(t_{X})+N_{V_{L^{*}-1}}(t_{X})\right).\end{align*}
 Simple manipulation gives us\begin{align*}
 & \text{ }N_{V_{L^{*}-1}}(t_{X})+N_{V_{L^{*}+1}}(t_{X})\\
\leq & \text{ }\frac{4\log n}{w^{\epsilon}}\left(N_{V_{L^{*}-2}}(t_{X})+N_{V_{L^{*}+2}}(t_{X})\right)\end{align*}
 with high probability. Proceeding with similar spirit gives us: for
all $s\left(0\leq s<L^{*}\leq\log n/2\right)$\begin{align}
 & \text{ }N_{V_{L^{*}-s}}(t_{X})+N_{V_{L^{*}+s}}(t_{X})\nonumber \\
\leq & \text{ }\frac{4\log n}{w^{\epsilon}}\left(N_{V_{L^{*}-s-1}}(t_{X})+N_{V_{L^{*}+s+1}}(t_{X})\right)\label{eq:NvDecayAdjacent}\end{align}
 holds with probability at least $1-n^{-4}$. By iteratively applying
(\ref{eq:NvDecayAdjacent}) we can derive that for any $s\text{ }\left(1\leq s<L^{*}\right)$,\begin{align*}
 & \text{ }N_{V_{s}}(t_{X})\\
\leq & \text{ }N_{V_{s}}(t_{X})+N_{V_{2L^{*}-s}}(t_{X})\\
\leq & \text{ }\left(\frac{4\log n}{w^{\epsilon}}\right)^{s-1}\left(N_{V_{1}}(t_{X})+N_{V_{2L^{*}-1}}(t_{X})\right)\\
\leq & \text{ }\left(\frac{4\log n}{w^{\epsilon}}\right)^{s-1}\left(c_{31}\sqrt{n}\log n\right),\end{align*}
 where the last inequality arises from the fact that there are at
most $O\left(\sqrt{n}\log n\right)$ nodes residing in each strip
with high probability.

This shows the geometric decaying rate of $N_{V_{s}}(t_{X})$ in $s$.
Suppose that $l^{*}>\log n/2$, then we have\begin{align*}
N_{V_{\frac{1}{2}\log n}}(t_{X}) & =\left(\frac{4\log n}{w^{\epsilon}}\right)^{\frac{1}{2}\log n-1}\left(c_{31}\sqrt{n}\log n\right)\\
 & \ll1\quad=\text{ }o\left(w^{-\epsilon}\log n\right),\end{align*}
 where contradiction arises. Hence, $l^{*}<\log n/2$ with high probability.

Additionally, we can derive an upper bound on $N_{V_{\log n/2}}$
using the same fixed-point arguments as follows\begin{align*}
 & \text{ }N_{V_{\log n/2}}(t_{X})\\
\leq & \text{ }\max\left\{ \log^{2}n,\right.\\
 & \quad\quad\left.\left(\frac{4\log n}{w^{\epsilon}}\right)^{\frac{1}{2}\log n-1}\left(N_{V_{1}}(t_{X})+N_{V_{\log n-1}}(t_{X})\right)\right\} \\
= & \text{ }\text{ }\log^{2}n\end{align*}
 with probability at least $1-n^{-3}$.

\subsection{Proof of Lemma \ref{Lemma-Cross-Strip-Remaining} \label{sec:Proof-of-Lemma-Cross-Strip-Remaining}}

Define a set of Bernoulli random variables $\left\{ X_{t}\right\} $
such that $X_{t}=1$ if there is at least one node inside these substrips
selecting $M^{*}$ for transmission at time $t$ and $X_{t}=0$ otherwise.
By observing that the size of each {}``substrip'' will not exceed
$\Theta\left(\log^{3}n\right)$ tiles before $T_{2}^{*}$ and that
each tile contains $O\left(\log n\right)$ nodes, the probability
$\mathbb{P}\left(X_{t}=1\right)$ can be bounded above by $1-\left(1-\frac{1}{w}\right)^{\Theta\left(\log^{3}n\right)}\leq\frac{c_{40}\log^{3}n}{w}$
for some constant $c_{40}$. This inspires us to construct the following
set of i.i.d. Bernoulli random variables $\left\{ \overline{X}_{t}\right\} $
through coupling as follows:\begin{align}
\begin{cases}
\text{if }X_{t} & =1,\text{ then \ensuremath{\overline{\ensuremath{X_{t}}}}}=1;\\
\text{if }X_{t} & =0,\text{ then \ensuremath{\overline{\ensuremath{X_{t}}}}}=\begin{cases}
1, & \text{w.p.}\frac{\frac{c_{40}\log^{3}n}{w}-\mathbb{P}\left(X_{t}=1\right)}{1-\mathbb{P}\left(X_{t}=1\right)},\\
0, & \text{otherwise}.\end{cases}\end{cases}\end{align}
That said, \begin{equation}
\overline{X_{t}}=\begin{cases}
1, & \text{ with probability }\frac{c_{40}\log^{3}n}{w};\\
0, & \text{ otherwise}.\end{cases}\end{equation}
 Our way for constructing $\left\{ \overline{X_{t}}\right\} $ implies
that \begin{equation}
\sum_{t=1}^{T_{2}^{*}}\overline{X_{t}}\geq\sum_{t=1}^{T_{2}^{*}}X_{t}=\frac{\log n}{2}.\end{equation}
 Additionally, large deviation bounds yields\begin{align}
 & \text{ }\mathbb{P}\left(\sum_{t=1}^{t_{X}}\overline{X_{t}}\geq\frac{\log n}{2}\right)\nonumber \\
\leq & \text{ }\mathbb{P}\left(\sum_{t=1}^{t_{X}}\left(\overline{X_{t}}-\mathbb{E}\left(\overline{X_{t}}\right)\right)\geq\frac{\log n}{2}-\frac{\log^{3}n}{w^{\epsilon}}\right)\nonumber \\
\leq & \text{ }\mathbb{P}\left(\sum_{t=1}^{t_{X}}\left(\overline{X_{t}}-\mathbb{E}\left(\overline{X_{t}}\right)\right)\geq\frac{\log n}{3}\right).\label{eq:LargeDeviationRemainingCrossStrip}\end{align}
 Define $\tilde{\beta}-1:=\frac{w^{\epsilon}\log n}{3c_{40}\log^{3}n}=\frac{w^{\epsilon}}{3c_{40}\log^{2}n}$,
\cite[Theorem A.1.12]{Alon2008} gives\begin{align}
(\ref{eq:LargeDeviationRemainingCrossStrip}) & \leq\left(\frac{e}{\tilde{\beta}}\right)^{\frac{\log n}{3}}\leq\frac{1}{n^{5}},\end{align}
 which further results in\begin{equation}
\mathbb{P}\left(T_{2}^{*}>t_{X}\right)\geq1-\frac{1}{n^{5}}.\end{equation}
 To conclude, the message $M^{*}$ is unable to cross $V_{1}^{B}$
by time $t_{X}=w^{1-\epsilon}$ with high probability.

\subsection{Proof of Lemma \ref{lemHittingTime} \label{sec:Proof-of-Lemma-Hitting-Time}}

Define two sets of random variables $\left\{ X_{ij}(t)\right\} $
and $\left\{ Y_{ij}(t)\right\} $ to represent the coordinates of
$Z_{ij}(t)$ in two dimensions, respectively, i.e. $Z_{ij}(t)=\left(X_{ij}(t),Y_{ij}(t)\right)$.
Therefore, for any $A\in\mathcal{A}_{\text{bd}}$, we can observe\begin{align*}
 & \mbox{\text{ }}\mathbb{P}_{0}\left(Z_{ij}(t)=A\right)\\
\leq & \mbox{\text{ }}\mathbb{P}_{0}\left(X_{ij}(t)=\pm\frac{\sqrt{m}}{2}\bigcup Y_{ij}(t)=\pm\frac{\sqrt{m}}{2}\right)\\
\leq & \mbox{\text{ }}4\mathbb{P}_{0}\left(X_{ij}(t)\geq\frac{\sqrt{m}}{2}\right).\end{align*}
 Besides, we notice that $\left|X_{ij}(t+1)-X_{ij}(t)\right|\leq2$
and $\mathbb{E}\left(X_{ij}(t+1)-X_{ij}(t)\right)=0$, then large
deviation results implies\begin{align}
\mathbb{P}_{0}\left(X_{ij}(t)\geq\frac{\sqrt{m}}{2}\right)\leq\exp\left(-\frac{6m}{c_{\text{h}}t}\right)\end{align}
 for some constant $c_{\text{h}}$. We thus derive $\mathbb{P}_{0}\left(X_{ij}(t)\geq\frac{\sqrt{m}}{2}\right)\leq n^{-6}$
for any $t<\frac{m}{c_{\text{h}}\log n}$, which leads to\begin{align*}
 & \mathbb{P}_{0}\left(T_{\text{hit}}<\frac{m}{c_{\text{h}}\log n}\right)\\
\leq & \mbox{\text{ }}\mathbb{P}_{0}\left(\exists t<\frac{m}{c_{\text{h}}\log n}\text{ and }\exists A\in\mathcal{A}_{\text{bd}}\text{ s.t. }Z_{ij}(t)=A\right)\\
\leq & \mbox{\text{ }}\frac{m}{c_{\text{h}}\log n}\cdot4\sqrt{m}\cdot4n^{-6}\leq\frac{1}{n^{4}}.\end{align*}

\subsection{Proof of Lemma \ref{lemSingleWalk}\label{sec:Proof-of-Lemma-SingleWalk}}

It can be observed that $Z_{ij}(t)$ is a discrete-time random walk
which at each step randomly moves to one of 25 sites each with some
constant probability. Formally, we can express it as follows: for
$\left|a\right|\leq2,\left|b\right|\leq2$: \begin{equation}
\mathbb{P}\left(Z_{ij}(t+1)-Z_{ij}(t)=(a,b)\mid Z_{ij}(t)\right)=p_{\left|a\right|,\left|b\right|}\end{equation}
 holds before $Z_{ij}(t)$ hits the boundary, where $p_{\left|a\right|,\left|b\right|}$
are fixed constants independent of $n$, and $\underset{\left|a\right|\leq2,\left|b\right|\leq2}{\sum}p_{\left|a\right|,\left|b\right|}=1$.
We construct a new process $\tilde{Z}_{ij}(t)$ such that $\tilde{Z}_{ij}(t)$
is a random walk over an {}``infinite'' plane with $\mathbb{P}\left(\tilde{Z}_{ij}(t+1)-\tilde{Z}_{ij}(t)=(a,b)\mid\tilde{Z}_{ij}(t)\right)=p_{\left|a\right|,\left|b\right|}.$

Define the event $\mathcal{H}_{\text{bd}}=\left\{ T_{\text{hit}}<m/\left(c_{\text{h}}\log n\right)\right\} $
where $T_{\text{hit}}$ is the hitting time to the boundary as defined
in Lemma \ref{lemHittingTime}. When restricted to the sample paths
where $Z_{ij}(t)$ does not hit the boundary by $t$, we can couple
the sample paths of $Z_{ij}(t)$ to those of $\tilde{Z}_{ij}(t)$
before the corresponding hitting time to the boundary.

It is well known that for a random walk $\tilde{Z}_{ij}(t)$ over
an infinite 2-dimensional plane, the return probability obeys $\tilde{q}_{ij}^{0}\sim t^{-1}$
\cite{FosterGood53}. Specifically, there exists a constant $\tilde{c}_{3}$
such that \begin{align}
\tilde{q}_{ij}^{0}(t) & \overset{\Delta}{=}\mathbb{P}_{0}\left(\tilde{Z}_{ij}(t)=(0,0)\right)\leq\tilde{c}_{3}/t.\end{align}
 Hence, the return probability $q_{ij}^{0}(t)\overset{\Delta}{=}\mathbb{P}_{0}\left(Z_{ij}(t)=(0,0)\right)$
of the original walk $Z_{ij}(t)$ satisfies\begin{align}
 & \text{ }q_{ij}^{0}(t)\nonumber \\
= & \text{ }\mathbb{P}_{0}\left(\left.Z_{ij}(t)=(0,0)\right|\mathcal{H}_{\text{bd}}\right)\mathbb{P}\left(\mathcal{H}_{\text{bd}}\right)\nonumber \\
 & \quad+\mathbb{P}_{0}\left(Z_{ij}(t)=(0,0)\wedge\overline{\mathcal{H}}_{\text{bd}}\right)\\
\leq & \text{ }\mathbb{P}_{0}\left(\mathcal{H}_{\text{bd}}\right)+\mathbb{P}_{0}\left(\tilde{Z}_{ij}(t)=(0,0)\wedge\overline{\tilde{\mathcal{H}}}_{\text{bd}}\right)\label{eq:CoupleInfiniteWalk}\\
\leq & \text{ }\mathbb{P}_{0}\left(\mathcal{H}_{\text{bd}}\right)+\mathbb{P}_{0}\left(\tilde{Z}_{ij}(t)=(0,0)\right)\nonumber \\
\leq & \text{ }\frac{1}{n^{4}}+\frac{\tilde{c}_{3}}{t}\quad\leq\frac{2\tilde{c}_{3}}{t},\nonumber \end{align}
 where (\ref{eq:CoupleInfiniteWalk}) arises from the coupling of
$Z_{ij}(t)$ and $\tilde{Z}_{ij}(t)$, and the upper bound on $\mathbb{P}\left(\mathcal{H}_{\text{bd}}\right)$
is derived in Lemma \ref{lemHittingTime}. Hence, the expected number
of time slots in which $i$ and $j$ move to the same subsquare by
time $t$$\left(1<t<\frac{m}{c_{h}\log n}\right)$ can be bounded
above as\begin{align*}
 & \mbox{\text{ }}\mathbb{E}\left(\left.\sum_{l=1}^{t}\mathds{1}\left(Z_{ij}(l)=(0,0)\right)\right|Z_{ij}(0)=(0,0)\right)\\
\leq & \mbox{\text{ }}2\tilde{c}_{3}\sum_{l=1}^{t}\frac{1}{l}\quad\leq\mbox{\text{ }}c_{3}\log t\end{align*}
 for some constant $c_{3}$.

\subsection{Proof of Lemma \ref{lemma-NumOfFailureBcRetransmissions}\label{sec:Proof-of-Lemma-Number-Of-Fairlure}}

Define $t_{i}(j)$ as the first time slot that $j$ receives $M_{i}$.
A {}``conflict'' event related to $M_{i}$ is said to occur if at
any time slot the source $i$ moves to a subsquare that coincides
with any user $j$ already possessing $M_{i}$. Denote by $C_{i}(t)$
the total amount of conflict events related to $M_{i}$ that happen
before $t$$\left(t\leq\frac{m}{c_{\text{h}}\log n}\right)$, which
can be characterized as follows\begin{align}
 & \mbox{\text{ }}\mathbb{E}\left(C_{i}(t)\right)\nonumber \\
=\mbox{\text{ }} & \mathbb{E}\left(\sum_{j\in\mathcal{N}_{i}(t)}\sum_{k=t_{i}(j)+1}^{t}\mathds{1}\left(Z_{ij}(k)=0\left|Z_{ij}\left(t_{i}(j)\right)=0\right.\right)\right)\\
\leq\mbox{\text{ }} & t\mathbb{E}\left(\sum_{k=1}^{t}\mathds{1}\left(Z_{ij}(k)=0\left|Z_{ij}(0)=0\right.\right)\right)\label{eq:ConflictLine2}\\
\leq\mbox{\text{ }} & c_{3}t\log t,\label{eq:ConflictLine3}\end{align}
 where (\ref{eq:ConflictLine2}) arises from the facts that $N_{i}\left(t\right)\leq t$
and $Z_{ij}$ is stationary, and (\ref{eq:ConflictLine3}) follows
from Lemma \ref{lemSingleWalk}. Lemma \ref{lemmaConcentration-3}
implies that there will be more than $n/6m$ users residing in each
subsquare with high probability. If this occurs for every subsquare
in each of $t$ slots, whenever $i$ and $j$ happens to stay in the
same subsquare, the probability that $i$ can successfully transmit
$M_{i}$ to $j$ can be bounded above as $\theta\left(1-\theta\right)\cdot6m/n$.
Therefore, the amount of successful {}``retransmissions'' can be
bounded as\begin{equation}
\mathbb{E}\left(F_{i}(t)\right)\leq\theta\left(1-\theta\right)\cdot\frac{6m}{n}\mathbb{E}\left(C_{i}(t)\right)\leq c_{5}\frac{mt\log t}{n}\end{equation}
 for some fixed constant $c_{5}$. Setting $t=t_{0}$ yields\begin{equation}
\mathbb{E}\left(F_{i}(t_{0})\right)\leq c_{5}\frac{m\log n}{n}t_{0}=o\left(t_{0}\right).\end{equation}
 This inequality is based on the assumption that $T_{\text{hit}}>t_{0}$
and that concentration effect occurs, which happens with probability
at least $1-3n^{-3}$.

\subsection{Proof of Lemma \ref{lemma-Phase1-Ni-Bound}\label{sec:Proof-of-Lemma-Phase1-Ni-Bound}}

Recall that $P_{1,j}$ denotes the $j^{\text{th}}$ subphase of Phase
1. Therefore, in the $1^{\text{st}}$ odd subphase, the total number
of successful transmissions from $i$ (including {}``retransmissions''),
denoted by $G_{i}$, can be bounded below as\begin{equation}
G_{i}\geq\tilde{c}_{7}m/\log n\label{eq:GiLowerBound}\end{equation}
 with probability exceeding $1-c_{7}/n^{4}$ for some constants and
$c_{7}$ and $\tilde{c}_{7}$. This follows from simple concentration
inequality and the fact that $\mathbb{E}\left(G_{i}\right)\geq c\left(1-\theta\right)\theta t$.
It can be noted that $G_{i}$ exhibits sharp concentration based on
simple large deviation argument. Also, Markov bound yields\begin{equation}
\mathbb{P}\left(F_{i}(t_{0})>65c_{5}\frac{m\log n}{n}t_{0}\right)\leq\frac{\mathbb{E}\left(F_{i}(t_{0})\right)}{65c_{5}mt_{0}\log n/n}=\frac{1}{65}.\end{equation}
 Since the duration of any odd subphase $t_{0}=m/\left(c_{\text{h}}\log n\right)$
obeys $mt_{0}\log n/n\ll t_{0}$, this implies that the number of
distinct new users that can receive $M_{i}$ within this odd subphase,
denoted by $\overline{N}_{i}\left(P_{1,1}\right)$, can be stochastically
bounded as\begin{equation}
\mathbb{P}\left(\overline{N}_{i}\left(P_{1,1}\right)<\tilde{c}_{8}\frac{m}{\log n}\right)<\frac{1}{64}\end{equation}
 for some constant $\tilde{c}_{8}$. Note that we use the bound $\frac{1}{64}$
here instead of $\frac{1}{65}$ in order to account for the deviation
of $G_{i}$.

We observe that there are in total $c_{6}m\log n/c_{\text{h}}$ slots
for spreading in Phase 1, therefore the total number of distinct nodes
receiving $M_{i}$ in Phase 1 is bounded above by $c_{6}m\log n/c_{\text{h}}\ll n/2$.
Mixing behavior combined with concentration effect thus indicates
that during any odd subphase, there are at least $n/12m$ nodes not
containing $M_{i}$ in each subsquare, and each transmitter is able
to contact a new receiver (who does not have $M_{i}$) in each slot
with probability at least $1/12$. This motivates us to construct
the following stochastically slower process for odd subphase $j$:
(a) there are $n/2$ nodes residing in the entire square; (b) each
transmission event is declared {}``failure'' regardless any other
state with probability $11/12$; (c) the evolution in different odd
subphases are independent; (d) other models and strategies remain
the same as the original process.

Obviously, this constructed process allows us to derive a lower bound
on $N_{i}(t)$ by the end of Phase 1. Proceeding with similar spirit
in the analysis for $1^{\text{st}}$ odd subphase, we can see that
the number of distinct new users receiving $M_{i}$ in odd subphase
$j$ of our constructed process, denoted by $\tilde{N}_{i}(P_{1,j})$,
can be lower bounded as\begin{equation}
\mathbb{P}\left(\tilde{N}_{i}\left(P_{1,j}\right)<\hat{c}_{8}\frac{m}{\log n}\right)<\frac{1}{64}\end{equation}
 for some constant $\hat{c}_{8}$. By noticing that $\left\{ \tilde{N}_{i}(P_{1,j})\right\} $
are mutually independent, simple large deviation inequality yields\begin{align*}
 & \mathbb{P}\left(\exists\frac{c_{6}}{2}\mbox{\ensuremath{\log}}^{2}n\text{ odd subphases s.t. }\tilde{N}_{i}\left(P_{1,j}\right)>\hat{c}_{8}\frac{m}{\log n}\right)\\
> & 1-\frac{1}{n^{5}}.\end{align*}
 Taking $c_{6}$ such that $c_{6}\hat{c}_{8}/2>32$, we can see that
\begin{equation}
\sum_{\text{odd subphase }j}\tilde{N}_{i}\left(P_{1,j}\right)>32m\log n\end{equation}
 holds with probability at least $1-o(n^{-3})$. Since $N_{i}\left(t\right)$
in the original process is stochastically larger than the total number
of distinct users receiving $M_{i}$ in any subphase $\sum_{\text{odd subphase }j}\tilde{N}_{i}\left(P_{1,j}\right)$,
we can immediately observe: $N_{i}\left(t\right)>32m\log n$ by the
end of Phase 1 with high probability. Furthermore, applying union
bound over all distinct messages completes the proof of this lemma.

\subsection{Proof of Lemma \ref{lemmaPhase2}\label{sec:Proof-of-Lemma-Phase2}}

We consider the evolution for a typical time block $[t_{B},t_{B}+1]$.
During any time slot in this time block, each node $l\notin\mathcal{\tilde{N}}_{i}^{B}(t_{B})$
lying in subsquare $A_{k}$ will receive $M_{i}$ from a node in $\mathcal{\tilde{N}}_{i}^{B}(t_{B})$
with probability at least\begin{align*}
 & \mbox{\text{ }}\mathbb{P}\left(l\text{ receives }M_{i}\text{ at }t\text{ from a node in }\mathcal{\tilde{N}}_{i}^{B}(t_{B})\right)\\
\geq\mbox{\text{ }} & c_{8}\frac{\tilde{N}_{i,A_{k}}(t)}{2n/m}\frac{1}{k}\geq\frac{c_{9}\tilde{N}_{i}^{B}(t_{B})}{nk}\end{align*}
 for constants $c_{8}$ and $c_{9}$, because concentration results
imply that: (1) there will be at least $\tilde{N}_{i}^{B}(t_{B})/3m$
nodes belonging to $\mathcal{\tilde{N}}_{i}^{B}(t_{B})$ residing
in each subsquare $A_{j}$; (2) there are at most $2n/m$ nodes in
each subsquare; (3) each successful transmission allows $l$ to receive
a specific message $M_{i}$ with state-independent probability $1/k$.
Therefore, each node $j\notin\mathcal{\tilde{N}}_{i}(t_{B})$ will
receive $M_{i}$ from a node in $\tilde{\mathcal{N}}_{i}^{B}(t_{B})$
by $t_{B}+1$ with probability exceeding $1-\left(1-\frac{c_{9}\tilde{N}_{i}^{B}(t_{B})}{nk}\right)^{k}\geq\tilde{c}_{9}\tilde{N}_{i}^{B}(t_{B})/n$
for some constant $\tilde{c}_{9}$. Since there are in total $n-\tilde{N}_{i}^{B}(t_{B})$
nodes not containing $M_{i}$ at the beginning of this time block,
we have \begin{align*}
 & \mathbb{P}\left(l\text{ receives }M_{i}\text{ from a node in }\mathcal{\tilde{N}}_{i}^{B}(t_{B})\text{ by }t_{B}+1\right)\\
\geq\mbox{\text{ }} & \left(n-\tilde{N}_{i}^{B}(t_{B})\right)\tilde{c}_{9}\tilde{N}_{i}^{B}(t_{B})/n\\
\geq\mbox{\text{ }} & \frac{\tilde{c}_{9}}{2}\tilde{N}_{i}^{B}(t_{B}).\end{align*}

Moreover, after the relaxation period of $c_{13}\log n/v^{2}(n)$
time slots, all these nodes containing $M_{i}$ will be spread out
to all subsquares, then similar arguments will hold for a new time
block. The mixing time period plays an important role in maintaining
an approximately uniform distribution of the locations of each node.
Thus, we can derive the following evolution equation:\begin{align}
 & \mathbb{E}\left(\left.\frac{1}{\tilde{N}_{i}^{B}(t_{B}+1)}-\frac{1}{\tilde{N}_{i}^{B}(t_{B})}\right|\tilde{N}_{i}^{B}(t_{B})\right)\\
= & -\mathbb{E}\left(\left.\frac{\tilde{N}_{i}^{B}(t_{B}+1)-\tilde{N}_{i}^{B}(t_{B})}{\tilde{N}_{i}^{B}(t_{B}+1)\tilde{N}_{i}^{B}(t_{B})}\right|\tilde{N}_{i}^{B}(t_{B})\right)\\
\leq & -\mathbb{E}\left(\left.\frac{\tilde{N}_{i}^{B}(t_{B}+1)-\tilde{N}_{i}^{B}(t_{B})}{2\left[\tilde{N}_{i}^{B}(t_{B})\right]^{2}}\right|\tilde{N}_{i}^{B}(t_{B})\right)\label{eq:Phase2NiStepE}\\
\leq & -c_{13}\frac{1}{\tilde{N}_{i}(t_{B})}\end{align}
 for some constant $c_{13}<1$, where the inequality (\ref{eq:Phase2NiStepE})
follows from the fact $\tilde{N}_{i}^{B}(t_{B}+1)\leq2\tilde{N}_{i}^{B}(t_{B})$
guaranteed by Step (4). Let $\tilde{T}_{i,2}^{B}=\min\left\{ t_{B}:\tilde{N}_{i}^{B}(t_{B})>n/8\right\} $,
and define $\tilde{Z}_{i}(t_{B})=\left(1-c_{13}\right)^{-t_{B}}/\tilde{N}_{i}^{B}(t_{B})$.
Simple manipulation yields\[
\mathbb{E}\left(\tilde{Z}_{i}\left((t_{B}+1)\wedge\tilde{T}_{i,2}^{B}\right)\mid\tilde{Z}_{i}\left(t_{B}\wedge\tilde{T}_{i,2}^{B}\right)\right)\leq\tilde{Z}_{i}\left(t_{B}\wedge\tilde{T}_{i,2}^{B}\right),\]
which indicates that $\left\{ \tilde{Z}_{i}\left(t_{B}\wedge\tilde{T}_{i,2}^{B}\right),t_{B}\geq0\right\} $
forms a non-negative supermartingale. We further define a stopping
time $\hat{T}_{i,2}^{B}=\tilde{T}_{i,2}^{B}\wedge n$, which satisfies
$\hat{T}_{i,2}^{B}\leq n<\infty$. The Stopping Time Theorem yields
\cite[Theorem 5.7.6]{Dur2010}\begin{align*}
\mathbb{E}\left(\tilde{Z}_{i}\left(\hat{T}_{i,2}^{B}\right)\right) & =\mathbb{E}\left(\tilde{Z}_{i}\left(\hat{T}_{i,2}^{B}\wedge\tilde{T}_{i,2}^{B}\right)\right)\\
 & \leq\mathbb{E}\left(\tilde{Z}_{i}\left(0\right)\right)=\frac{1}{\tilde{N}_{i}^{B}(0)},\end{align*}
\begin{equation}
\Longrightarrow\quad\mathbb{E}\left(\left(1-c_{13}\right)^{-\hat{T}_{i,2}^{B}}\right)\leq\frac{\max\tilde{N}_{i}^{B}\left(\hat{T}_{i,2}^{B}\right)}{\tilde{N}_{i}^{B}(0)}<n.\end{equation}
 Set $c_{14}=(1-c_{13})^{-1}$, then we have\[
\mathbb{P}\left\{ \hat{T}_{i,2}^{B}>4\log_{c_{14}}n\right\} =\mathbb{P}\left\{ c_{14}^{\hat{T}_{i,2}^{B}}>n^{4}\right\} \leq\frac{\mathbb{E}\left(c_{14}^{\hat{T}_{i,2}^{B}}\right)}{n^{4}}<\frac{1}{n^{3}}.\]
The fact that $\hat{T}_{i,2}^{B}=\tilde{T}_{i,2}^{B}\wedge n$ gives\[
\mathbb{P}\left\{ \tilde{T}_{i,2}^{B}\leq4\log_{c_{14}}n\right\} =\mathbb{P}\left\{ \hat{T}_{i,2}^{B}\leq4\log_{c_{14}}n\right\} \geq1-\frac{1}{n^{3}}.\]
Finally, applying union bound yields \[
\mathbb{P}\left\{ \exists i:\tilde{T}_{i,2}^{B}>4\log_{c_{14}}n\right\} \leq\frac{1}{n^{2}}.\]

\subsection{Proof of Lemma \ref{lemmaPhase3Unicast}\label{sec:Proof-of-Lemma-Phase3-Unicast}}

Define $T_{i,3}=\min\left\{ t:N_{i}(t)=n\right\} $ as the stopping
time for all $i$. In each time slot, any node $l\notin\mathcal{N}_{i}(t)$
will receive $M_{i}$ with probability at least $c_{18}/16n$ for
a constant $c_{18}$ simply because each transmitter in $\mathcal{N}_{i}(t)$
will select $M_{i}$ for transmission to $l$ with probability at
least $1/k$. This yields the following inequality:\begin{equation}
\mathbb{E}\left(N_{i}(t+1)-N_{i}(t)\left|N_{i}(t)\right.\right)\geq\frac{c_{18}}{16k}\left(n-N_{i}(t)\right).\label{eq:StayMovePhase4}\end{equation}
 Define $Z_{i,3}(t)=\left(n-N_{i}(t)\right)\left(1-\frac{c_{18}}{16k}\right)^{-t}$,
then by manipulation we get $\mathbb{E}\left(Z_{i,3}(t+1)\mid Z_{i,3}(t)\right)\geq Z_{i,3}(t),$
which indicates that $\left\{ Z_{i,3}(t)\right\} $ forms a submartingale.
Take $t_{4}=64k\log n/c_{18}$, we have\[
\mathbb{E}\left[n-N_{i}(t_{4})\right]\leq\left(1-\frac{c_{18}}{16k}\right)^{t_{4}}\frac{15n}{16},\]
\[
\therefore\quad\mathbb{P}\left(N_{i}\left(t_{4}\right)\leq n-1\right)\leq\mathbb{E}\left[\left(n-N_{i}(t_{4})\right)\right]\leq\frac{15}{16n^{3}}.\]
 Moreover, applying union bound yields\[
\mathbb{P}\left(\exists i:N_{i}\left(t_{4}\right)\leq n-1\right)\leq n\mathbb{P}\left(N_{i}\left(t_{4}\right)\leq n-1\right)\leq\frac{15}{16n^{2}}.\]
 Let $T_{3}=\max_{i}T_{i,3}$. Combining the above results, we can
see that the duration $T_{3}$ of Phase $3$ satisfies $T_{3}=O\left(k\log n\right)$
with probability at least $1-\frac{15}{16n^{2}}$.

{}

{}

\bibliographystyle{IEEEtran} \bibliographystyle{IEEEtran} \bibliographystyle{IEEEtran}
\bibliographystyle{IEEEtran} \bibliographystyle{IEEEtran} \bibliographystyle{IEEEtran}
\bibliographystyle{IEEEtran} \bibliographystyle{IEEEtran}
\bibliography{bibfileGossip}

\begin{thebibliography}{10}
\providecommand{\url}[1]{#1}
\csname url@samestyle\endcsname
\providecommand{\newblock}{\relax}
\providecommand{\bibinfo}[2]{#2}
\providecommand{\BIBentrySTDinterwordspacing}{\spaceskip=0pt\relax}
\providecommand{\BIBentryALTinterwordstretchfactor}{4}
\providecommand{\BIBentryALTinterwordspacing}{\spaceskip=\fontdimen2\font plus
\BIBentryALTinterwordstretchfactor\fontdimen3\font minus
  \fontdimen4\font\relax}
\providecommand{\BIBforeignlanguage}[2]{{%
\expandafter\ifx\csname l@#1\endcsname\relax
\typeout{** WARNING: IEEEtran.bst: No hyphenation pattern has been}%
\typeout{** loaded for the language `#1'. Using the pattern for}%
\typeout{** the default language instead.}%
\else
\language=\csname l@#1\endcsname
\fi
#2}}
\providecommand{\BIBdecl}{\relax}
\BIBdecl

\bibitem{ChenInfocom}
Y.~Chen, S.~Shakkottai, and J.~Andrews, ``Sharing multiple messages over mobile
  networks,'' \emph{IEEE Infocom}, pp. 658 --666, April 2011.

\bibitem{QiuSri04}
D.~Qiu and R.~Srikant, ``Modeling and performance analysis of bittorrent-like
  peer-to-peer networks,'' \emph{ACM SIGCOMM}, pp. 367--378, 2004.

\bibitem{YanDeV04}
X.~Yang and G.~de~Veciana, ``Service capacity of peer to peer networks,''
  \emph{IEEE INFOCOM}, vol.~4, pp. 2242 -- 2252 vol.4, March 2004.

\bibitem{Bor87}
B.~Pittel, ``On spreading a rumor,'' \emph{SIAM Journal of Applied
  Mathematics}, vol.~47, no.~1, pp. 213--223, 1987.

\bibitem{KarSchSheVoc00}
R.~Karp, C.~Schindelhauer, S.~Shenker, and B.~Vocking, ``Randomized rumor
  spreading,'' \emph{Proceedings of the 41st Annual Symposium on Foundations of
  Computer Science}, pp. 565--574, 2000.

\bibitem{Tsi84}
J.~Tsitsiklis, \emph{Problems in decentralized decision making and
  computation}, PhD dissertation, LIDS, MIT, Cambridge, 1984.

\bibitem{KasSri07}
A.~Kashyap and R.~Srikant, ``Quantized consensus,'' \emph{Automatica}, vol.~43,
  pp. 1192--1203, 2007.

\bibitem{BoydShah06}
S.~Boyd, A.~Ghosh, B.~Prabhakar, and D.~Shah, ``Randomized gossip algorithms,''
  \emph{IEEE Transactions on Information Theory}, vol.~52, no.~6, pp.
  2508--2530, June 2006.

\bibitem{NedOzd09}
A.~Nedic and A.~Ozdaglar, ``Distributed subgradient methods for multi-agent
  optimization,'' \emph{IEEE Transactions on Automatic Control}, vol.~54,
  no.~1, pp. 48 --61, Jan. 2009.

\bibitem{JunShaShi10}
K.~Jung, D.~Shah, and J.~Shin, ``Distributed averaging via lifted markov
  chains,'' \emph{IEEE Transactions on Information Theory}, vol.~56, no.~1, pp.
  634 --647, January 2010.

\bibitem{ModShaZus}
E.~Modiano, D.~Shah, and G.~Zussman, ``Maximizing throughput in wireless
  networks via gossiping,'' \emph{ACM SIGMETRICS}, pp. 27--38, 2006.

\bibitem{EryOzdShaMod10}
A.~Eryilmaz, A.~Ozdaglar, D.~Shah, and E.~Modiano, ``Distributed cross-layer
  algorithms for the optimal control of multihop wireless networks,''
  \emph{IEEE/ACM Transactions on Networking}, vol.~18, no.~2, pp. 638--651,
  Apr. 2010.

\bibitem{Dem88}
A.~Demers, D.~Greene, C.~Houser, W.~Irish, J.~Larson, S.~Shenker, H.~Sturgis,
  D.~Swinehart, and D.~Terry, ``Epidemic algorithms for replicated database
  maintenance,'' \emph{Proceeding of the sixth annual {ACM} symposium on
  principles of distributed computing}, pp. 1--12, 1987.

\bibitem{Pittel1987}
B.~Pittel, ``On spreading a rumor,'' \emph{SIAM Journal on Applied
  Mathematics}, vol.~47, no.~1, pp. pp. 213--223, 1987.

\bibitem{GossipTutorialShah}
D.~Shah, ``Gossip algorithms,'' \emph{Foundations and Trends in Networking},
  vol.~3, no.~1, pp. 1--125, April 2009.

\bibitem{SanHajMas07}
S.~Sanghavi, B.~Hajek, and L.~Massoulie, ``Gossiping with multiple messages,''
  \emph{IEEE Transactions on Information Theory}, vol.~53, no.~12, pp.
  4640--4654, Dec. 2007.

\bibitem{MoskSha08}
D.~Mosk-Aoyama and D.~Shah, ``Fast distributed algorithms for computing
  separable functions,'' \emph{IEEE Transactions on Information Theory},
  vol.~54, no.~7, pp. 2997--3007, July 2008.

\bibitem{MobilityGossip}
A.~Sarwate and A.~Dimakis, ``The impact of mobility on gossip algorithms,''
  \emph{IEEE INFOCOM}, pp. 2088--2096, April 2009.

\bibitem{Clementi2011}
A.~Clementi, A.~Monti, F.~Pasquale, and R.~Silvestri, ``Information spreading
  in stationary {M}arkovian evolving graphs,'' \emph{IEEE Transactions on
  Parallel and Distributed Systems}, vol.~22, no.~9, pp. 1425 --1432, Sep.
  2011.

\bibitem{Baumann2009}
H.~Baumann, P.~Crescenzi, and P.~Fraigniaud, ``Parsimonious flooding in dynamic
  graphs,'' \emph{Proceedings of the 28th ACM symposium on Principles of
  distributed computing}, pp. 260--269, 2009.

\bibitem{Clementi2012}
\BIBentryALTinterwordspacing
A.~Clementi, R.~Silvestri, and L.~Trevisan, ``Information spreading in dynamic
  graphs,'' Nov. 2011. [Online]. Available:
  \url{http://arxiv.org/abs/1111.0583}
\BIBentrySTDinterwordspacing

\bibitem{Pettarin2011}
A.~Pettarin, A.~Pietracaprina, G.~Pucci, and E.~Upfal, ``Tight bounds on
  information dissemination in sparse mobile networks,'' \emph{Proceedings of
  the 30th annual ACM SIGACT-SIGOPS symposium on principles of distributed
  computing}, pp. 355--362, 2011.

\bibitem{DebMedCho06}
S.~Deb, M.~Medard, and C.~Choute, ``Algebraic gossip: a network coding approach
  to optimal multiple rumor mongering,'' \emph{IEEE Transactions on Information
  Theory}, vol.~52, no.~6, pp. 2486--2507, June 2006.

\bibitem{Hae2011}
B.~Haeupler, ``Analyzing network coding gossip made easy,'' \emph{43rd ACM
  Symposium on Theory of Computing}, pp. 2088--2096, June 2011.

\bibitem{Zhe2006}
R.~Zheng, ``Information dissemination in power-constrained wireless networks,''
  \emph{IEEE INFOCOM}, pp. 1 --10, April 2006.

\bibitem{SubShaAra}
S.~Subramanian, S.~Shakkottai, and A.~Arapostathis, ``Broadcasting in sensor
  networks: The role of local information,'' \emph{IEEE/ACM Transactions on
  Networking}, vol.~16, no.~5, pp. 1133 --1146, Oct. 2008.

\bibitem{ResSan2010}
G.~Resta and P.~Santi, ``On the fundamental limits of broadcasting in wireless
  mobile networks,'' \emph{IEEE INFOCOM}, pp. 1 --5, 14-19 2010.

\bibitem{NeeMod05}
M.~Neely and E.~Modiano, ``Capacity and delay tradeoffs for ad hoc mobile
  networks,'' \emph{IEEE Transactions on Information Theory}, vol.~51, no.~6,
  pp. 1917--1937, June 2005.

\bibitem{MobilityCapacityTse}
M.~Grossglauser and D.~Tse, ``Mobility increases the capacity of ad hoc
  wireless networks,'' \emph{IEEE/ACM Transactions on Networking}, vol.~10,
  no.~4, pp. 477--486, Aug 2002.

\bibitem{GamMamProSha06}
A.~El~Gamal, J.~Mammen, B.~Prabhakar, and D.~Shah, ``Optimal throughput-delay
  scaling in wireless networks -- {Part I}: the fluid model,'' \emph{IEEE
  Transactions on Information Theory}, vol.~52, no.~6, pp. 2568 --2592, June
  2006.

\bibitem{YinYanSri08}
L.~Ying, S.~Yang, and R.~Srikant, ``Optimal delay-throughput tradeoffs in
  mobile ad hoc networks,'' \emph{IEEE Transactions on Information Theory},
  vol.~54, no.~9, pp. 4119 --4143, Sep. 2008.

\bibitem{LinMazShr06}
X.~Lin, G.~Sharma, R.~Mazumdar, and N.~Shroff, ``Degenerate delay-capacity
  tradeoffs in ad-hoc networks with {B}rownian mobility,'' \emph{IEEE
  Transactions on Information Theory}, vol.~52, no.~6, pp. 2777 --2784, June
  2006.

\bibitem{ChenGunes07}
C.~Avin and G.~Ercal, ``On the cover time and mixing time of random geometric
  graphs,'' \emph{Theoretical Computer Science}, vol. 380, no. 1-2, pp. 2--22,
  2007.

\bibitem{RandomGraphDynamics}
R.~Durrett, \emph{Random Graph Dynamics}, ser. Cambridge Series in Statistical
  and Probabilistic Mathematics.\hskip 1em plus 0.5em minus 0.4em\relax New
  York, NY: Cambridge University Press.

\bibitem{Alon2008}
N.~Alon and J.~H. Spencer, \emph{The Probabilistic Method}, ser.
  Wiley-Interscience Series in Discrete Mathematics and Optimization.\hskip 1em
  plus 0.5em minus 0.4em\relax John Wiley \& Sons, Inc., 2008.

\bibitem{FosterGood53}
F.~G. Foster and I.~J. Good, ``On a generalization of {P}\'{o}lya's random-walk
  theorem,'' \emph{The Quarterly Journal of Mathematics}, vol.~4, pp. 120--126,
  June 1953.

\bibitem{Dur2010}
R.~Durrett, \emph{Probability: Theory and Examples (4th Edition)}.\hskip 1em
  plus 0.5em minus 0.4em\relax Cambridge University Press, 2010.

\end{thebibliography}

\end{document}